\documentclass{aa}

\pdfoutput=1

\usepackage{graphicx}
\usepackage{txfonts}
\usepackage[usenames,dvipsnames]{color}

\usepackage[breaklinks=true]{hyperref} %% to avoid \citeads line fills

\newcommand{\Sref}[1]{Section \ref{#1}}
\newcommand{\Tref}[1]{Table \ref{#1}}
\newcommand{\Aref}[1]{Appendix \ref{#1}}
\sfcode`\.=1002\sfcode`\?=1002\sfcode`\!=1002
\newcommand{\Fref}[1]{\ifhmode \ifnum\spacefactor=1002 Figure \ref{#1}\else Fig.\ \ref{#1}\fi \else Figure \ref{#1}\fi}
\newcommand{\Eref}[1]{\ifhmode \ifnum\spacefactor=1002 Equation (\ref{#1})\else equation (\ref{#1})\fi \else Equation (\ref{#1})\fi}

\newcommand{\tran}[3]{\ensuremath{\ion{#1}{#2}\,\lambda\textrm{#3}}}

\newcommand{\kms}{\ensuremath{\textrm{km\,s}^{-1}}}
\newcommand{\SN}{\ensuremath{\textrm{S/N}}}
\newcommand{\chisq}{\ensuremath{\chi^2}}

\newcommand{\zem}{\ensuremath{z_\text{\scriptsize em}}}
\newcommand{\zab}{\ensuremath{z_\text{\scriptsize abs}}}

\newcommand{\daa}{\ensuremath{\Delta\alpha/\alpha}}
\newcommand{\cms}{\ensuremath{\textrm{cm\,s}^{-1}}}
\newcommand{\ms}{\ensuremath{\textrm{m\,s}^{-1}}}

\begin{document} 

\title{Fundamental physics with ESPRESSO: Precise limit on variations in the fine-structure constant towards the bright quasar HE\,0515$-$4414\thanks{Based on Guaranteed Time Observations collected at the European Southern Observatory under ESO programme 1102.A-0852 by the ESPRESSO Consortium.}, \thanks{The data and analysis products from this work are publicly available at \href{https://doi.org/10.5281/zenodo.5512490}{https://doi.org/10.5281/zenodo.5512490} \citep{Murphy_2021Spec+Fits_HE0515}.}}
%\subtitle{I. Overviewing the $\kappa$-mechanism}
\titlerunning{ESPRESSO $\alpha$ measurement with quasar HE\,0515$-$4414}

\author{
Michael T. Murphy\inst{1,2}\and
Paolo Molaro\inst{3,2}\and
Ana C. O. Leite\inst{4,5,6}\and
Guido Cupani\inst{3,2}\and
Stefano Cristiani\inst{3,2}\and
Valentina D'Odorico\inst{3,2}\and
Ricardo G\'enova Santos\inst{7,8}\and
Carlos J. A. P. Martins\inst{4,6}\and
Dinko Milakovi\'c\inst{3,2,9}\and
Nelson J. Nunes\inst{10,11}\and
Tobias M. Schmidt\inst{12,3}\and
Francesco A. Pepe\inst{12}\and
Rafael Rebolo\inst{7,8}\and
Nuno C. Santos\inst{4,5}\and
S\'ergio G. Sousa\inst{4,5}\and
Maria-Rosa Zapatero Osorio\inst{13}\and
Manuel Amate\inst{7}\and
Vardan Adibekyan\inst{4,5}\and
Yann Alibert\inst{14}\and
Carlos Allende Prieto\inst{7,8}\and
Veronica Baldini\inst{3}\and
Willy Benz\inst{14}\and
Fran\c{c}ois Bouchy\inst{12}\and 
Alexandre Cabral\inst{10,11}\and 
Hans Dekker\inst{15}\and
Paolo Di Marcantonio\inst{3}\and
David Ehrenreich\inst{12}\and
Pedro Figueira\inst{4,16}\and
Jonay I. Gonz\'alez Hern\'andez\inst{7,8}\and
Marco Landoni\inst{17}\and
Christophe Lovis\inst{12}\and
Gaspare Lo Curto\inst{16}\and
Antonio Manescau\inst{15}\and
Denis M\'egevand\inst{12}\and
Andrea Mehner\inst{16}\and
Giuseppina Micela\inst{18}\and
Luca Pasquini\inst{15}\and
Ennio Poretti\inst{17,19}\and
Marco Riva\inst{17}\and
Alessandro Sozzetti\inst{20}\and
Alejandro Su\'arez Mascare\~no\inst{7,8}\and
St\'ephane Udry\inst{4}\and
Filippo Zerbi\inst{17}
}

\authorrunning{M. T. Murphy et al.}

\institute{
Centre for Astrophysics and Supercomputing, Swinburne University of Technology, Hawthorn, Victoria 3122, Australia,
\email{mmurphy@swin.edu.au}\and
Institute for Fundamental Physics of the Universe, Via Beirut 2, I-34151 Miramare, Trieste, Italy\and  
INAF -- Osservatorio Astronomico di Trieste, via G. B. Tiepolo 11, I-34143 Trieste, Italy\and
Instituto de Astrof\'isica e Ci\^encias do Espa\c co, CAUP, Universidade do Porto, Rua das Estrelas, 4150-762, Porto, Portugal\and  
Departamento de F\'isica e Astronomia, Faculdade de Ci\^encias, Universidade do Porto, Rua Campo Alegre, 4169-007, Porto, Portugal\and  
Centro de Astrof\'{\i}sica da Universidade do Porto, Rua das Estrelas, 4150-762 Porto, Portugal\and  
Instituto de Astrof\'{\i}sica de Canarias (IAC), Calle V\'{\i}a L\'actea s/n, E-38205 La Laguna, Tenerife, Spain\and 
Departamento de Astrof\'{\i}sica, Universidad de La Laguna (ULL), E-38206 La Laguna, Tenerife, Spain\and 
Istituto Nazionale di Fisica Nucleare, Sezione di Trieste, Via Bonomea 265, 34136 Trieste, Italy\and
Instituto de Astrof\'isica e Ci\^encias do Espa\c{c}o, Faculdade de Ci\^encias da Universidade de Lisboa, Campo Grande, PT1749-016 Lisboa, Portugal\and
Departamento de Física da Faculdade de Ciências da Universidade de Lisboa, Edifício C8, 1749-016 Lisboa, Portugal\and
Observatoire Astronomique de l’Universit\'e de Gen\`eve, Chemin Pegasi 51, Sauverny 1290, Switzerland\and
Centro de Astrobiolog\'{\i}a (CSIC-INTA), Crta. Ajalvir km 4, E-28850 Torrej\'on de Ardoz, Madrid, Spain\and 
Physics Institute, University of Bern, Sidlerstrasse 5, 3012 Bern, Switzerland\and
European Southern Observatory, Karl-Schwarzschild-Strasse 2, 85748, Garching b. M\"unchen, Germany\and 
European Southern Observatory, Alonso de Co\'ordova 3107, Vitacura, Regio\'on Metropolitana, Chile\and 
INAF -- Osservatorio Astronomico di Brera, Via E. Bianchi 46, I-23807 Merate, Italy\and
INAF -- Osservatorio Astronomico di Palermo, Piazza del Parlamento 1, I-90134 Palermo, Italy\and 
INAF -- Osservatorio Nazionale Galileo Rambla José Ana Férnandez Pérez, 7 38712 San Antonio de Breña Baja Spain 
INAF -- Osservatorio Astrofisico di Torino, via Osservatorio 20, 10025 Pino Torinese, Italy
}

\date{Received September 19, 2021; accepted XXX XX, 20XX}

% \abstract{}{}{}{}{} 
% 5 {} token are mandatory
\abstract{
The strong intervening absorption system at redshift 1.15 towards the very bright quasar HE\,0515$-$4414 is the most studied absorber for measuring possible cosmological variations in the fine-structure constant, $\alpha$. We observed HE\,0515$-$4414 for 16.1\,h with the Very Large Telescope and present here the first constraint on relative variations in $\alpha$ with parts-per-million (ppm) precision from the new ESPRESSO spectrograph: $\daa = 1.3 \pm 1.3_{\rm stat} \pm 0.4_{\rm sys}\,{\rm ppm}$. The statistical uncertainty (1$\sigma$) is similar to the ensemble precision of previous large samples of absorbers, and derives from the high \SN\ achieved ($\approx$105 per 0.4\,\kms\ pixel). ESPRESSO's design, and calibration of our observations with its laser frequency comb, effectively removed wavelength calibration errors from our measurement. The high resolving power of our ESPRESSO spectrum ($R=145000$) enabled the identification of very narrow components within the absorption profile, allowing a more robust analysis of \daa. The evidence for the narrow components is corroborated by their correspondence with previously detected molecular hydrogen and neutral carbon. The main remaining systematic errors arise from ambiguities in the absorption profile modelling, effects from redispersing the individual quasar exposures, and convergence of the parameter estimation algorithm. All analyses of the spectrum, including systematic error estimates, were initially blinded to avoid human biases. We make our reduced ESPRESSO spectrum of HE\,0515$-$4414 publicly available for further analysis. Combining our ESPRESSO result with 28 measurements, from other spectrographs, in which wavelength calibration errors have been mitigated, yields a weighted mean $\daa = -0.5 \pm 0.5_{\rm stat} \pm 0.4_{\rm sys}$\,ppm at redshifts 0.6--2.4.
}

\keywords{quasars: absorption lines -- quasars: individual: \object{HE 0515$-$4414} -- intergalactic medium -- cosmology: observations -- cosmology: miscellaneous -- instrumentation: spectrographs\vspace{-4em}}

\maketitle

\section{Introduction}\label{s:intro}

The standard model of particle physics is characterised by a set of
physical interactions, whose strengths are described through
fundamental dimensionless couplings. Historically, these have been
assumed to be spacetime-invariant. However, fundamental couplings are
known to run with energy, and in many extensions of the standard model
they also change in time and possibly in space -- see the recent
review of this field by \citet{Martins_2017RPPh...80l6902M}.

The fine-structure constant, $\alpha \equiv e^{2}/\hbar c$, and the
proton-to-electron mass ratio, $\mu$, are two dimensionless
fundamental constants that can be probed directly with spectroscopic
techniques. In Earth-based laboratories, comparing atomic clocks based
on different transition frequencies over $\sim$2--4\,yr timescales has
provided extraordinarily precise limits on local time variations
\citep[e.g.][]{Rosenband_2008Sci...319.1808R}. The most recent
measurements constrain the time rate of change in $\alpha$ and $\mu$
to just $(1.0 \pm 1.1)\times10^{-18}$ and
$(-8 \pm 36)\times10^{-18}\,\mathrm{yr}^{-1}$, respectively
\citep{Lange_2021PhRvL.126a1102L}. While it may be tempting to
linearly extrapolate such constraints to cosmological time (and
distance) scales, it should be emphasised that how the fundamental
constants may vary, and on what they may depend, is entirely unknown;
such extrapolations would be just as simplistic as assuming no
variation at all. Clearly, the variability or constancy of $\alpha$
and $\mu$ must be explicitly tested over the full range of time (and
distance) scales available to experiments.

The most effective way to probe cosmological variations in $\alpha$ is
by measuring any relative shifts between metal transitions produced in
intervening quasar absorption systems. This technique was introduced
by \citet{Dzuba_1999PhRvA..59..230D,Dzuba_1999PhRvL..82..888D} and
\citet{Webb_1999PhRvL..82..884W} and is known as the Many Multiplet
(MM) method because it utilises the wavelengths of many different
transitions from several atomic species. Transitions from different
multiplets have different dependencies on $\alpha$, so measuring the
velocity shifts between these transitions in a quasar absorption
system provides a direct probe of \daa\ -- the relative difference
between $\alpha$ in the absorber ($\alpha_{\rm abs}$) and its current
laboratory value on Earth ($\alpha_{\rm lab}$):
\begin{equation}
\frac{\Delta \alpha}{\alpha} \equiv \frac{\alpha_{\rm abs}-\alpha_{\rm lab}}{\alpha_{\rm lab}} \approx -\frac{\Delta v_i}{c}\frac{1}{2Q_i},
\label{e:alpha}
\end{equation}
where $\Delta v_i$ is the velocity shift caused by a small variation
in $\alpha$ \citep[i.e.\ $\daa \ll
1$;][]{Dzuba_2002PhRvA..66b2501D}. $Q_i \equiv \omega_i/q_i$, where
$q_i$ is the sensitivity coefficient: the expected sensitivity of
transition $i$'s laboratory rest wavenumber ($\omega_i$) to variations
in $\alpha$. These $q$ coefficients have been calculated in numerous
works, using several different many-body quantum mechanical
techniques, starting with \citet{Dzuba_1999PhRvA..59..230D} and
compiled in \citet{Murphy_2014MNRAS.438..388M}. For typical values of
$q$, a 1\,ppm variation in $\alpha$ would produce a shift of order
$\Delta v\sim 20$\,\ms\ between different transitions.

The MM method has been widely used to measure \daa\ at large
redshifts, i.e.\ large look-back-times, with the largest samples
obtained with archival spectra from two high-resolution spectrographs:
UVES at the VLT and HIRES at the Keck
Observatory. \citet{Webb_2011PhRvL.107s1101W} and
\citet{King_2012MNRAS.422.3370K} combined the sample of 143 Keck/HIRES
absorption systems in
\citet{Murphy_2003MNRAS.345..609M,Murphy_2004LNP...648..131M} with a
sample of 154 from VLT/UVES to produce a data set of
\daa\ measurements in 293 distinct absorption systems. This sample
showed a statistical preference for dipolar spatial variation of
$\alpha$ across the sky at the $\sim$10\,ppm level with $>$4$\sigma$
statistical significance \citep{Webb_2011PhRvL.107s1101W}. As should
be expected for such a surprising result, with potential impact on
fundamental physics, many authors have questioned the data, analysis,
assumptions and potential systematic errors underpinning these
measurements, and presented alternative data sets and analyses
\citep[e.g.][]{Chand_2004A&A...417..853C,Quast_2004A&A...415L...7Q,Levshakov_2005A&A...434..827L,Levshakov_2007A&A...466.1077L,Molaro_2008EPJST.163..173M}. However,
these did not provide compelling alternative explanations or results
\citep[e.g.][]{Murphy_2007PhRvL..99w9001M,Murphy_2008MNRAS.384.1053M,Wilczynska_2015MNRAS.454.3082W}. Constraints
from higher quality spectra of individual absorbers were also
obtained, but none of them directly supported or strongly conflicted
with the $\alpha$ dipole evidence
\citep[e.g.][]{Molaro_2013A&A...555A..68M,Bainbridge_2017MNRAS.468.1639B,Wilczynska_2020SciA....6.9672W}.

It should be emphasised that a spatial variation is a theoretically
unexpected result. While many well-motivated models allow redshift
dependencies of $\alpha$, possibly with additional dependencies on the
local density \citep[e.g.][]{Martins_2017RPPh...80l6902M}, a
theoretical model which can account for a pure spatial dipole is more
difficult to identify and requires significant fine-tuning
\citep{Olive_2011PhRvD..83d3509O}. Testing this result with improved
\emph{accuracy} -- i.e.\ decreased systematic errors -- is therefore
essential.

One possibility is that the evidence for an $\alpha$ dipole may
originate from different -- possibly similar -- systematic errors in
the two spectrographs (Keck/HIRES and VLT/UVES). By observing solar
spectra reflected from asteroids with UVES, and comparing them with a
more accurately wavelength calibrated solar spectrum from a Fourier
Transform Spectrometer, \citet{Rahmani_2013MNRAS.435..861R} found that
the usual thorium--argon (ThAr) lamp calibration produced long-range
wavelength distortions. Gradients in these distortions were as large
as $\sim$400\,\ms\ over 600\,\AA\ ranges. Using 20 years of archival
HIRES and UVES spectra of asteroids and solar twins,
\citet{Whitmore_2015MNRAS.447..446W} found that long-range distortions
in their (ThAr-based) wavelength scales were ubiquitous and
substantial, with gradients typically varying between
$\pm$200\,\ms\,per 1000\,\AA, potentially causing spurious shifts in
\daa\ at the $\sim$10\,ppm level. Using a simple model of the
distortions, they were able to substantially explain the previous
evidence for $\alpha$ variations from the large UVES sample, and
partially explain that of the HIRES sample. The smaller samples of HIRES and
UVES spectra cited above would be similarly affected.

While it remains plausible that more detailed distortion modelling may
not fully explain the HIRES and UVES results
\citep{Dumont_2017MNRAS.468.1568D}, the additional uncertainties
introduced into the \daa\ measurements by such models would greatly
decrease the $\alpha$ dipole's statistical significance. Overall, the
wavelength calibration distortions in slit-based spectrographs
substantially undermine confidence in the evidence for spatial
variations in $\alpha$. Furthermore, recent quasar observations with
UVES, HIRES and Subaru/HDS, dedicated to measuring $\alpha$ and
explicitly corrected for distortions (with contemporaneous solar twin
and asteroid spectra), or insensitive to them (utilising transitions
close together in wavelength), do not support spatial variations but
are only inconsistent with the dipole at the 2$\sigma$ level
\citep{Evans_2014MNRAS.445..128E,Murphy_2016MNRAS.461.2461M,Kotus_2017MNRAS.464.3679K,Murphy_2017MNRAS.471.4930M}. On
balance, there is currently no compelling evidence for variations in
$\alpha$ over cosmological time or distance scales. Nevertheless, even
when accurate corrections for distortions are possible, the
uncertainty in the wavelength calibration of slit-based spectrographs
dominates the total error budget.

The new high-resolution spectrograph at the VLT, ESPRESSO
\citep[Echelle SPectrograph for Rocky Exoplanet and Stable
Spectroscopic Observations;][]{Pepe_2021A&A...645A..96P}, was
specifically designed to suppress wavelength calibration errors in
quasar absorption measurements of $\alpha$
\citep{Molaro_2006IAUS..232..198M,Molaro_2009ASSP....9..389M}. ESPRESSO
is fed by optical fibres, is sealed in a stable vacuum vessel with
temperature control at the mK level, and can be calibrated with an
`astrocomb' -- a femtosecond-pulsed laser frequency comb (LFC) with
secondary mode filtering adapted to ESPRESSO's resolving power. In
principle, the high frequency-space density of uniformly separated
(filtered) comb modes, whose individual frequencies are known
\textit{a priori} to $\ll$1\,\cms\ accuracy, could enable
$\sim$1\,\cms\ (photon-limited) calibration precision
\citep{Murphy_2007MNRAS.380..839M}. The ESPRESSO astrocomb
calibration, described in detail by
\citet{Schmidt_2021A&A...646A.144S} and Lovis et al.\ (in prep.) and
summarised below (\Sref{ss:wavecal}), covers the wavelength range
$\sim$4850--7000\,\AA\ (cf.\ ESPRESSO's full range of 3780--7780\,\AA)
and provides a $\la$1\,\ms\ accuracy for velocities measured between
the transitions in the spectrum analysed here
(\Sref{ss:wavecal}). This is well below the ensemble statistical
precision of the quasar absorption measurements ($\sim$30\,\ms), so
ESPRESSO effectively removes wavelength calibration from the error
budget. The fibre feed and highly stabilised environment ensures that
drifts in ESPRESSO's wavelength calibration are far too slow ($>$days)
and small ($\la$1\,\ms) to affect quasar absorption measurements of
$\alpha$.

Here we make the first precise ESPRESSO measurement of \daa, with LFC
calibration providing a highly accurate wavelength scale. We selected
HE\,0515$-$4414 as a high-value target because it is one of the
brightest quasars in the southern sky, with an intervening absorption
system at redshift $\zab=1.1508$ which places the most important metal
transitions within the LFC-calibrated wavelength range. The absorption
profile also contains many separate, relatively narrow features which
appear in many different transitions; in simple terms, this increases
the number of centroid measurements that combine to constrain
\daa. These advantages mean this absorber has been studied numerous
times and provided the most precise cosmological constraint on \daa\
from any individual system
\citep[][]{Quast_2004A&A...415L...7Q,Levshakov_2006A&A...449..879L,Chand_2006A&A...451...45C,Molaro_2008EPJST.163..173M,Kotus_2017MNRAS.464.3679K,Milakovic_2021MNRAS.500....1M}.

\section{Data acquisition and reduction}\label{s:data}

\subsection{Observations}\label{ss:obs}

HE\,0515$-$4414 (J2000 right ascension
05$^\textrm{h}$17$^\textrm{m}$07.6$^\textrm{s}$ and declination
$-44^\circ$10$^\prime$55.58$^{\prime\prime}$) was identified as a very
bright ($m_{v}\approx 15.2$\,mag; \textsl{Gaia} $G=14.90$\,mag) quasar
at redshift $\zem=1.71$ by \citet{Reimers_1998A&A...334...96R}. It was
part of the ESPRESSO Consortium's Guaranteed Time Observations (GTO)
in two main runs: a visitor-mode run on 4--7 November 2018, and
service mode observations between November 2019 and March 2020.
\Tref{t:obs} provides a journal of the observations. The total
integration of 57916\,s was obtained over 17 exposures: 32400\,s over
9 exposures in 2018, and 25516\,s over 8 exposures in 2019 and
2020. ESPRESSO is located in the incoherent combined Coud\'e facility,
underneath the four Unit Telescopes (UTs), and can be fed by any UT
or, indeed, all four simultaneously. The high-resolution, single-UT
mode with 2-pixel binning in the spatial direction (i.e.\
`singleHR21') was selected for observing HE\,0515$-$4414, proving a
nominal resolving power of $R\sim145000$ with a 1\farcs0 diameter
fibre (in the red arm). All exposures were obtained with UT3-Melipal,
except for those in November 2019 which were observed with UT1-Antu. It was
later found that the atmospheric dispersion corrector of UT3 was not
perfectly aligned during the 2018 run, causing a $\sim$30\% loss in
throughput. Also, the fibre link was upgraded in June 2019 and this
increased the total efficiency by $\sim$40\%
\citep{Pepe_2021A&A...645A..96P}. The cloud and seeing conditions
varied considerably during our two runs. These factors combined to
produce variations in the signal-to-noise ratio (\SN) from $\approx$19
to 30 per $\approx$0.4\,\kms\ pixel at 6000\,\AA\ in the extracted
spectra. One exposure was terminated early, with $<$1/3 of the nominal
1\,h exposure time, producing a \SN\ of only $\approx$8.

\begin{table}
\caption{Journal of the observations under the ESPRESSO Consortium's GTO program (ESO Project ID 1102.A-0852). The airmass is the average value during the exposure, and IQ is the full-width-half-maximum image quality measured by the image analysis detector at the telescope's focus; this has been found empirically to correspond well with occasional image quality measurements taken with a detector at the end of the Coud\'e train. The signal-to-noise ratio (S/N) per 0.4\,\kms\ is measured near 6000\,\AA\ as an average over 4\,\AA\ in the extracted spectrum  (after combining both traces of the same echelle order).}
\label{t:obs}
\vspace{-1.5em}
\begin{center}
\begin{tabular}{ccccc}   
\hline\hline 
Exposure start   & Exposure & Airmass & IQ        & S/N  \\
(UTC)            & (s)      &         & ($^{\prime\prime}$) & \\
\hline
2018-11-04 07:00 & 3600     & 1.07    & 0.68      & 31 \\
2018-11-05 03:36 & 3600     & 1.33    & 1.03      & 20 \\
2018-11-05 04:37 & 3600     & 1.17    & 0.89      & 24 \\
2018-11-05 05:38 & 3600     & 1.09    & 0.84      & 27 \\
2018-11-05 06:39 & 3600     & 1.07    & 0.87      & 28 \\
2018-11-05 07:40 & 3600     & 1.10    & 0.82      & 28 \\
2018-11-07 05:21 & 3600     & 1.10    & 0.84      & 26 \\
2018-11-07 06:22 & 3600     & 1.07    & 1.00      & 26 \\
2018-11-07 07:27 & 3600     & 1.10    & 0.86      & 21 \\
2019-11-26 06:53 & 3500     & 1.15    & 0.92      & 30 \\
2020-02-28 01:14 & 1016     & 1.16    & 1.11      &  8 \\
2020-02-28 01:33 & 3500     & 1.26    & 0.86      & 25 \\
2020-02-28 02:34 & 3500     & 1.49    & 1.12      & 19 \\
2020-03-03 00:34 & 3600     & 1.16    & 1.16      & 22 \\
2020-03-19 02:34 & 3400     & 2.13    & 1.05      & 25 \\
2020-03-20 01:41 & 3500     & 1.68    & 1.14      & 26 \\
2020-03-21 01:39 & 3500     & 1.69    & 1.08      & 26 \\
\hline
\end{tabular}
\end{center}
\end{table}

\subsection{Data reduction and wavelength calibration}\label{ss:wavecal}

The data were reduced with version 2.2.3 of the standard ESPRESSO data
reduction software (DRS), based on ESO's \textsc{reflex}
environment\footnote{http://www.eso.org/pipelines}. The reduction
process is detailed in \citet{Pepe_2021A&A...645A..96P}. The
DRS-extracted spectra are supplied in
\citet{Murphy_2021Spec+Fits_HE0515} online. Most important for this
work is the LFC wavelength calibration of the quasar spectra; below we
focus on this aspect of the data reduction.

In the 2018 visitor-mode run, the LFC spectra were taken at the
beginning of each night, i.e.\ each quasar spectrum was calibrated
with a LFC exposure taken up to $\sim$9\,hr before. For the single
2019 service-mode exposure, the LFC calibration was taken $\sim$16\,hr
prior. In the 2020 service-mode run, three LFC exposures were taken on
27 and 29 February and 3 March, bracketing the first four quasar
exposures, but none were taken for the last three quasar exposures
(19--21 March) due to technical problems with the LFC. Those problems
may have affected the latter two LFC exposures. Indeed, a comparison
between the wavelength solutions from the three LFC exposures showed
differences of $\approx$3\,\ms\ in the blue arm's calibration, while
those in the red arm were $<$1\,\ms. We therefore calibrated all the
2020 quasar exposures with the 27 February LFC exposure.

In all science exposures, the quasar was observed in Fibre A while
Fibre B was used to record simultaneous sky spectra instead of
recording a simultaneous wavelength calibration spectrum. Therefore,
no information about possible instrumental drifts during a night is
available. However, ESPRESSO is highly stabilised, and drifts are
known to be less than 1\,\ms\ over a night
\citep{Pepe_2021A&A...645A..96P}. While it appears unlikely that the
changes in the last three LFC exposures above are due to instrument
drifts over the 1-week timescale, we cannot rule it out entirely. If
we assume those drifts are real and extrapolate them over the full
3-week timescale of the 2020 service-mode run, they would create
$\sim$9\,\ms\ differences between exposures in the blue arm and, more
importantly, a $\sim$5\,\ms\ relative shift between the two arms. The
latter could potentially cause a systematic, $\sim$0.3\,ppm shift in
\daa, though this would be significantly diluted by the
better-calibrated 2018 exposures. However, few of the transitions in
the $\zab=1.1508$ absorber of HE\,0515$-$4414 fall in the blue arm,
reducing this effect's impact on \daa. Indeed, we explicitly test the
effect of removing the blue-arm transitions from our analysis in
\Sref{sss:arms} and find it to be negligible. Note that this only
applies to the analysis of this specific absorber, with our specific
set of observations, and should not be generalised in future analyses
to mean that possible instrument drifts are never important. They
should be checked in each specific analysis of \daa\ with ESPRESSO.

The LFC wavelength calibration procedure is explained in
\citet{Schmidt_2021A&A...646A.144S} and Lovis et al.\ (in
prep.). Briefly, the LFC spectrum is imaged onto part of both ESPRESSO
CCDs, covering the wavelength range $\approx$4700--7200\,\AA\ (the
CCDs' wavelength ranges slightly overlap at
$\approx$5200--5270\,\AA). Given the drop in LFC intensity at both edges
of the spectrum, a satisfactory calibration can be obtained in a
slightly more restricted range from $\approx$4850--7000\,\AA. Outside
this range the wavelength solution is normally obtained from the
daytime standard ThAr and Fabry--P\'erot combined calibrations, which
are treated in a similar way to those used by HARPS
\citep{Cersullo_2019A&A...624A.122C}. The ESPRESSO DRS identifies the
LFC lines automatically: the LFC mode frequencies are known \textit{a
  priori} to very high accuracy, but the mode number of a single mode
must first be established, and this is achieved using a `rough' first
guess at the wavelength scale established with a traditional ThAr
lamp.

The LFC spectrum itself exhibits a significant background light
contribution and also strong modulation of the line intensities which
may originate in the Fabry--P\'erot mode-filtering cavities
\citep{Milakovic_2020MNRAS.493.3997M,Schmidt_2021A&A...646A.144S}. These
can be effectively removed with detailed modelling and subtraction of
the background light contribution
\citep{Schmidt_2021A&A...646A.144S}. Finally,
\citeauthor{Schmidt_2021A&A...646A.144S} identified an unexpected and,
so far, unexplained high-frequency pattern in the LFC calibration
residuals: the best-fit centroid wavelength for neighbouring LFC modes
alternate either side of the final wavelength solution by
$\approx$5--7\,\ms. However, for the absorption system studied here,
the individual transitions span $\approx$750\,\kms, or $\sim$80 LFC
modes, so any high-frequency residuals would average down to $\la$1\,\ms\ per
transition. Averaged over the $>$7 transitions that contribute the
strongest constraints on \daa, the systematic error should be
$<$0.5\,\ms, or $\la$0.02\,ppm in \daa, which is negligible.

\subsection{Combined spectrum}\label{ss:combspec}
   
\textsc{uves\_popler} \citep[version
1.05;][]{Murphy_2016zndo.....44765M,Murphy_2019MNRAS.482.3458M} was
used to combine the spectra extracted by the DRS to form a single
spectrum for subsequent analysis. The S/N of the combined spectrum is
$\approx$105\,per 0.4\,\kms\ pixel at 6000\,\AA, near the
\ion{Mg}{ii}\,2796/2803 doublet of the $\zab=1.1508$ absorber, and
$\approx$85 at 5000\,\AA\ near \ion{Fe}{ii} $\lambda$2344, the bluest
line that contributes strong constraints on \daa. It is one of the
highest quality echelle quasar spectra available: its \SN\ per \kms\
is $\approx$170 at 6000\,\AA, compared with $\approx$230 for the UVES
spectrum of the same quasar in the UVES SQUAD database
\citep{Murphy_2019MNRAS.482.3458M}, but with a substantially higher
resolving power: $R \approx 145000$ for ESPRESSO compared to 75000 for
the UVES spectrum.

There are two traces of each echelle order, so the 17 exposures
(\Tref{t:obs}) provide at least 34 independent flux measurements to be
considered for each pixel in the combined spectrum (68 for the
overlapping regions of neighbouring orders). This allows for
aggressive rejection of outlying flux values and a reliable final
(inverse-variance) weighted mean flux value. Upward spikes in flux
(`cosmic rays') are automatically flagged by \textsc{uves\_popler} on
each extracted order trace, and discrepant values are further flagged
via iterative $\sigma$-clipping during the combination
\citep{Murphy_2019MNRAS.482.3458M}; for the ESPRESSO spectrum of
HE\,0515$-$4414, individual values more than 2.5$\sigma$ above or
below the weighted mean were rejected. However, \textsc{uves\_popler}
is also designed for detailed visual inspection and comparison of
individual echelle order spectra, and this revealed many instances of
low-level `cosmic rays' and/or `hot pixels' that were not flagged by
the techniques above. These features were typically spread over 3--10
pixels; while most individual pixels escaped the automatic clipping
approaches above, the broader features were clearly visible during
manual inspection. The ESPRESSO DRS has very recently been improved,
and \textsc{uves\_popler} will be improved in future, to remove most
of these effects automatically. Nevertheless, for the present work,
they were removed based on visual inspection only. The detailed record
of these manual actions is contained within the \textsc{uves\_popler}
Log (UPL) file in \citet{Murphy_2021Spec+Fits_HE0515} online.

We also masked all relevant telluric emission and absorption features
around the transitions of interest for the $\zab=1.1508$ absorber,
which left small gaps in coverage and narrower regions of lower S/N,
in the \ion{Fe}{ii} $\lambda$2344 and 2586 lines, the \ion{Mg}{ii}
$\lambda\lambda$2796/2803 doublet and \ion{Mg}{i} $\lambda$2852. The
mask was created using a synthetic spectrum of the atmosphere for Cerro
Paranal with the default, recommended input parameter values in ESO's
{\sc skymodel} software\footnote{See
  http://www.eso.org/observing/etc/skycalc/skycalc.htm.}
\citep{Noll_2012A&A...543A..92N,Jones_2013A&A...560A..91J}. The
resolution was matched to our ESPRESSO observations (nominal $R=145000$) and a
unit airmass was assumed. Pixels in the synthetic spectrum whose
transmission was reduced by more than 0.5\% by telluric features,
relative to the local maximum value (within $\pm$200\,\kms), were
flagged. Given the S/N of the combined HE\,0515$-$4414 spectrum
($\sim$100\,per 0.4\,\kms\ pixel), this ensures all features potentially causing
a $>$1$\sigma$ suppression of flux, integrated over 5 pixels, will be
removed, assuming the features are aligned in all exposures. In
practice, the 17 different exposures have somewhat different
barycentric corrections, and so the Earth-frame telluric features are
shifted relative to each other by up to 17\,\kms\ between the 2018,
2019 and 2020 runs. Emission features were flagged, in a similar way,
as pixels with $>$2 times the background emission in the synthetic
spectrum than the minimum in the surrounding $\pm$200\,\kms. This
threshold was derived by visually inspecting the sky-subtraction
residuals near very strong emission lines, and so should be very
conservative. The mask was defined by excluding wavelength ranges
within 3\,\kms\ of flagged pixels, and applied to each exposure in
\textsc{uves\_popler} after the combination step.

\textsc{uves\_popler} automatically constructs a nominal continuum for
the combined spectrum using low-order polynomial fits. However, given
the very broad absorption features of the $\zab=1.1508$ system, we
manually re-fitted the continuum across each transition of
interest. Again, this is a manual and visually guided process, but the
log of these steps is transparently recorded in the UPL file, supplied
in \citet{Murphy_2021Spec+Fits_HE0515} online, which can easily be
used to fully reproduce the combined spectrum from the DRS-extracted
spectra \citep{Murphy_2019MNRAS.482.3458M}.

\section{Analysis}\label{s:analysis}

This section presents the full details of the analysis procedure. Casual readers may wish to skip the intricacies of the absorption profile modelling in Sections \ref{ss:model_right}--\ref{ss:model_left}, which will be more important for those seeking to reproduce our analysis approach and for comparison with other techniques.

\subsection{Comparison between ESPRESSO, HARPS and UVES spectra}\label{ss:spec_comp}

\begin{figure*}
\begin{center}
\includegraphics[width=0.99\textwidth]{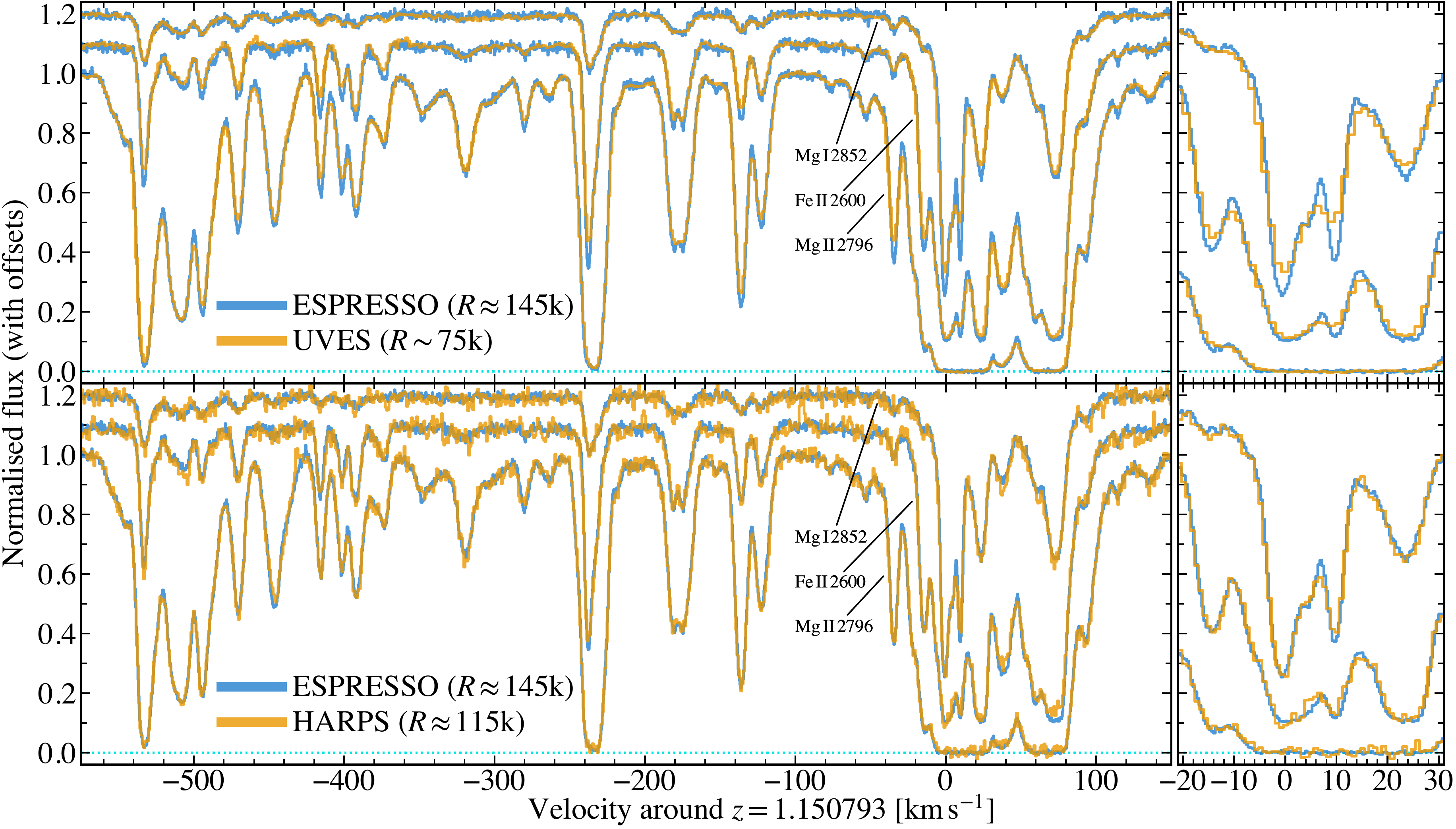} 
\vspace{-0.5em}
\caption{Comparison between continuum-normalised ESPRESSO, HARPS and UVES spectra of the $\zab=1.1508$ absorber towards HE\,0515$-$4414. The upper panels compare the ESPRESSO (blue) and UVES (orange) spectra, while the lower panels compare the ESPRESSO (blue) and HARPS (orange) spectra. In each panel, the upper, middle and lower spectra correspond to the three different transitions indicated -- \tran{Mg}{i}{2852}, \tran{Fe}{ii}{2600} and \tran{Mg}{ii}{2796}, respectively -- offset by 10\% in normalised flux for clarity. Note that the resolving powers ($R$) of the spectra are substantially different, but the differences between the spectra are minor, except for the features near 0 and 9\,\kms\ (see text). The right-hand panels focus on this region, where the absorption is strongest (note the different velocity scale).}
\label{f:spec_comp}
\end{center}
\end{figure*}

\Fref{f:spec_comp} compares the new, combined ESPRESSO spectrum of HE\,0515$-$4414 with those observed with HARPS \citep{Milakovic_2021MNRAS.500....1M} and UVES \citep{Kotus_2017MNRAS.464.3679K} in three strong transitions of the $\zab=1.1508$ absorber. The resolving powers of these three spectra are substantially different -- $R \sim 75000$ for UVES, $\approx$115000 for HARPS, and $\approx$145000 for ESPRESSO -- yet only small differences are apparent between the UVES and ESPRESSO spectra, and there is almost no discernable difference between the ESPRESSO and HARPS spectra. As expected, the largest differences between the ESPRESSO and UVES spectra appear in the cores of the sharpest, narrowest spectral features, but even those differences are not apparent between the ESPRESSO and HARPS spectra. This demonstrates that almost all the velocity structure of the $\zab=1.1508$ absorber is already resolved at $R \sim 100000$. This was anticipated by \citet{Kotus_2017MNRAS.464.3679K}: even the lower resolution UVES spectrum required a large number of closely-spaced (relative to their $b$-parameters) velocity components to fit the spectral features of the velocity structure, so increasing the resolving power was predicted to reveal little unresolved structure.

However, two exceptions to the general conclusion above are the spectral features near 0 and 9\,\kms\ in \Fref{f:spec_comp}. When comparing the spectra in the unsaturated \tran{Mg}{i}{2852} transition of these features, the difference between the ESPRESSO and UVES spectra is larger than for other features; the 9\,\kms\ feature even appears different in the ESPRESSO and HARPS spectra. This indicates the presence of particularly narrow (i.e.\ low $b$-parameter) velocity components near 0 and 9\,\kms. These correspond in redshift to the narrow \ion{C}{i} absorption lines tracing very cold gas detailed by \citet{Quast_2002A&A...386..796Q}, and are consistent with hosting the H$_2$ absorption discovered by \citet{Reimers_2003A&A...410..785R}. We discuss these features in \Sref{ss:model_right} when constructing our model fit to this region of the absorber.

\subsection{Absorption systems and transitions}\label{ss:systems}

\begin{figure*}
\begin{center}
\includegraphics[width=0.80\textwidth]{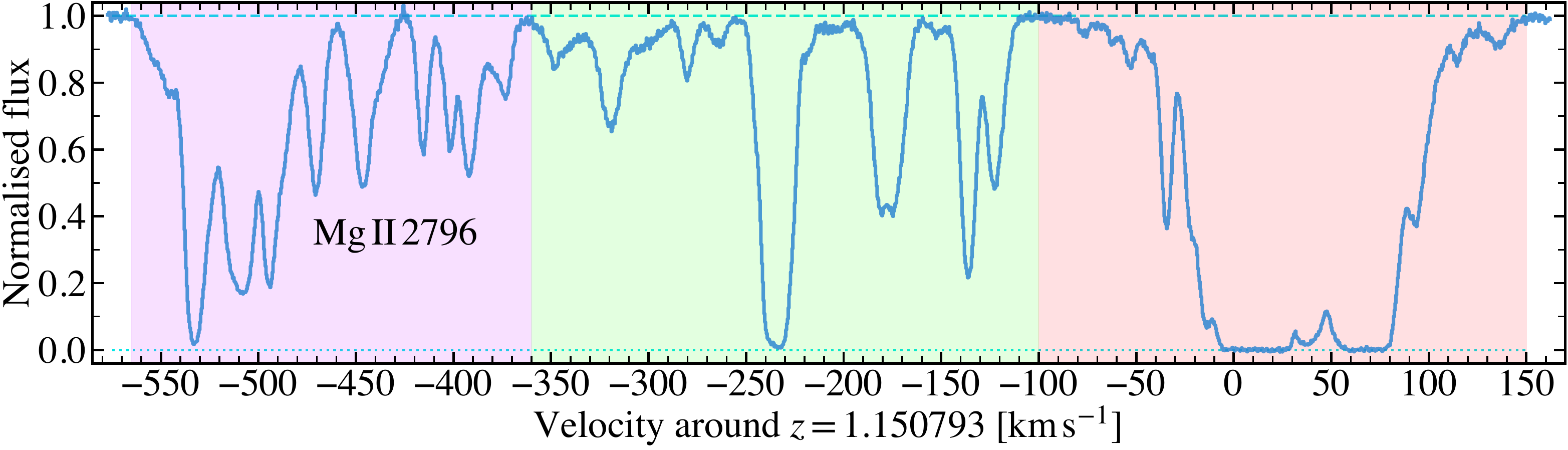} 
\vspace{-0.5em}
\caption{Definition of `regions' in the $\zab=1.1508$ absorber: left (purple shading; $-565$ to $-360$\,\kms), central (green; $-360$ to $-100$\,\kms) and right (red; $-100$ to 150\,\kms). The continuum-normalised flux in the strongest transition detected (\tran{Mg}{ii}{2796}) is plotted to illustrate that the flux returns to the fitted quasar continuum level at the region edges. While not strictly necessary in the fitting process, this helps avoid edge effects in the model when it is convolved with a (Gaussian) model of the ESPRESSO instrumental profile.}
\label{f:regions}
\end{center}
\end{figure*}

Given its very complex absorption profile, we simplified the analysis of the $\zab=1.1508$ absorber -- especially the profile fitting aspect -- by following \citet{Kotus_2017MNRAS.464.3679K}'s approach of splitting the velocity structure into three `regions', as illustrated in \Fref{f:regions}, defined in velocity relative to $\zab=1.150793$: left ($-565$ to $-360$\,\kms), central ($-360$ to $-100$\,\kms) and right ($-100$ to 150\,\kms). The strongest absorption and narrowest features are in the right region, so it is not surprising that all previous measurements in this absorber have found the strongest constraints on \daa\ from this part of the absorber. Therefore, while we analysed all three regions in the same level of detail, we describe the analysis of the right region more comprehensively here, and pay greater attention to systematic errors due to the smaller statistical uncertainty on \daa\ it provides.

\begin{figure}
\begin{center}
\includegraphics[width=\columnwidth]{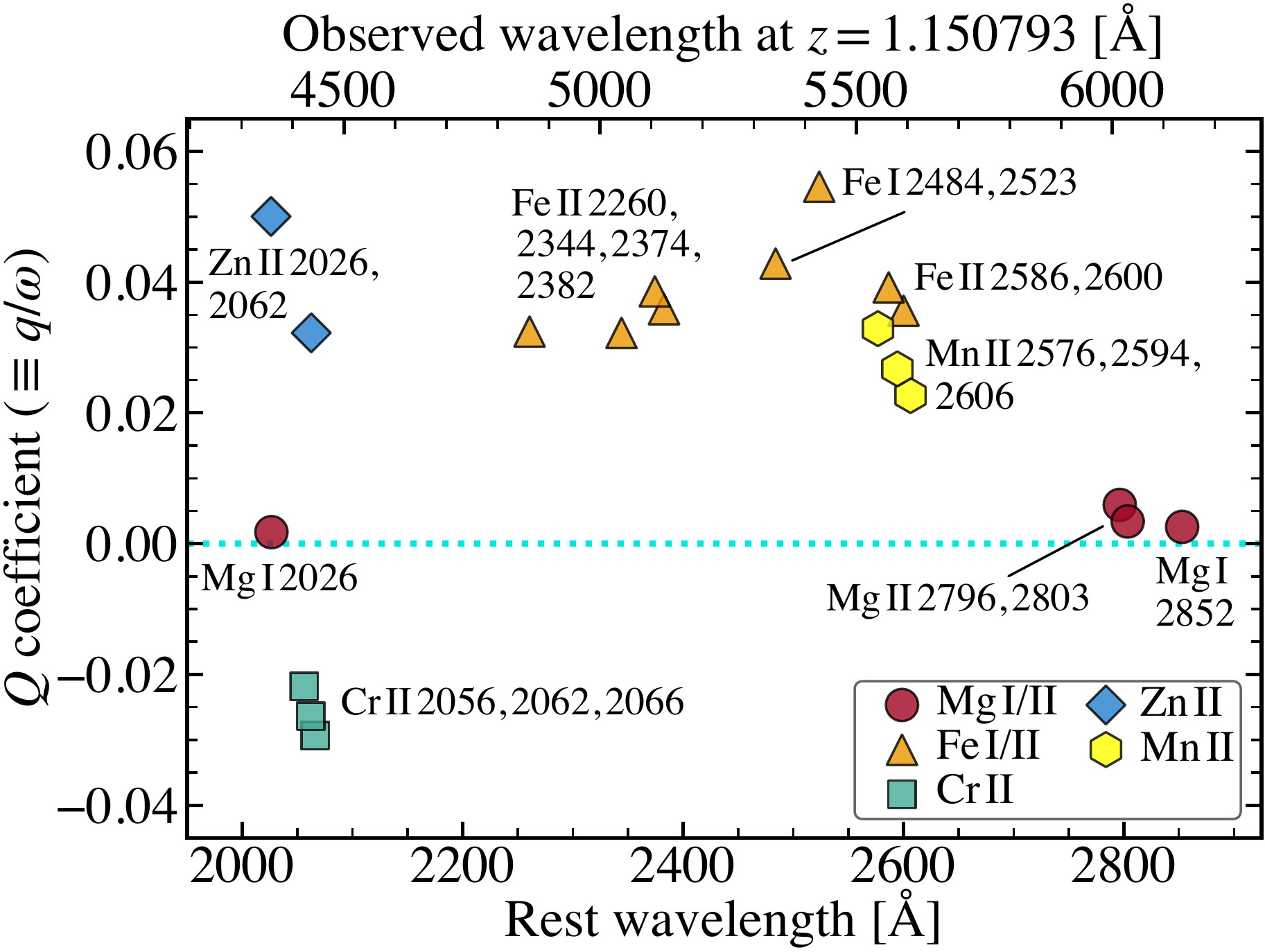}
\vspace{-1em}
\caption{Sensitivity ($Q$) coefficients for the transitions used here to constrain \daa\ in the $\zab=1.1508$ absorber towards HE\,0515$-$4414. The $Q$ coefficients were compiled in \citet{Murphy_2014MNRAS.438..388M} and are defined in \Eref{e:alpha}. The lower horizontal axis shows the laboratory rest wavelengths of the transitions, while the upper axis shows the observed wavelength at \zab.}
\label{f:Qvswl}
\end{center}
\end{figure}

\Fref{f:Qvswl} shows the transitions detected in our ESPRESSO spectrum and used to measure \daa\ in the $\zab=1.1508$ absorber. The plot shows the sensitivity coefficient, $Q$, of each transition, as defined in \Eref{e:alpha}. The main principle of the Many Multiplet method is to compare as many transitions, with as great a contrast of $Q$ coefficients, as possible. In practice, to understand the relative importance of each transition for constraining \daa, the $Q$ distribution in \Fref{f:Qvswl} has to be considered together with the absorption strength in each transition (including any saturation), and the narrowness of the features in the different parts of the velocity structure. The strongest available transitions in the $\zab=1.1508$ absorber are the \ion{Mg}{ii} doublet and the five \ion{Fe}{ii} lines at 2344--2600\,\AA. As \Fref{f:Qvswl} shows, this combination of transitions offers a high contrast in $Q$. Indeed, in the left and central regions, \Fref{f:spec_comp} shows that these transitions offer many narrow, unsaturated features to be compared. However, for the right region, the features that are strongest in \ion{Fe}{ii} are saturated in \ion{Mg}{ii}, so the \ion{Mg}{ii} doublet is less important. Instead, the \tran{Mg}{i}{2852} line offers unsaturated features and a very different $Q$ value to the \ion{Fe}{ii} lines; it is these 6 transitions that most strongly constrain \daa\ in the right region. Note that uncertainties in the $Q$ values, which are $\la$7\% for the \ion{Fe}{ii} transitions \citep{Murphy_2014MNRAS.438..388M}, will cause \daa\ to be systematically scaled from its true value by, at most, a similar proportion; these errors are reduced when using many transitions and are negligible for the purposes of this work in any case. Also note that the form of \Eref{e:alpha} means that systematic errors in the $Q$ values cannot generate a spurious, non-zero \daa. Finally, as indicated in \Fref{f:Qvswl}, many other transitions are detected, from \ion{Mg}{i}, \ion{Fe}{i}, \ion{Zn}{ii}, \ion{Cr}{ii} and \ion{Mn}{ii}. These are all very weak, even in the right region, and do not constrain \daa\ strongly. However, fitting the weak \tran{Mg}{i}{2026} transition simultaneously with the much stronger \tran{Mg}{i}{2852} line has important consequences for our model fit to the right region, as discussed in detail below (\Sref{ss:model_right}).

While we focus on measuring \daa\ in the $\zab=1.1508$ absorber in this work, it is important to identify other intervening absorbers in case any of their transitions blend with those of the $z=1.1508$ system. Given its relatively narrow wavelength range (3780--7780\,\AA), the ESPRESSO spectrum is not ideal for this purpose: clearly identifying any absorber between $\zab=0$ and $\zab=\zem=1.71$ requires a large wavelength range in which to confidently detect at least two transitions. By contrast, the wavelength coverage of the UVES spectrum of \citet{Kotus_2017MNRAS.464.3679K} is much wider (3051--10430\,\AA) and better suited for this. Apart from the $\zab=1.1508$ absorber and Galactic absorption, \citeauthor{Kotus_2017MNRAS.464.3679K} identified 8 others in the UVES spectrum at redshifts $\zab = 0.2223$, 0.2818, 0.4291, 0.9406, 1.3849, 1.5145, 1.6737 and 1.6971. For the transitions we fit, the only known blend is with the \tran{Fe}{ii}{2344} transition from \tran{Ca}{ii}{3934} in the $\zab=0.2818$ absorber \citep{Kotus_2017MNRAS.464.3679K}. We therefore do not fit this \ion{Fe}{ii} transition between $-4.5$ and 4.5\,\kms\ or redwards of 80\,\kms\ in the right region. However, during our profile fitting analysis of the right and central regions below (Sections \ref{ss:model_right} and \ref{ss:model_central}) we detected additional absorption in two transitions (\tran{Mg}{ii}{2803} and \tran{Fe}{ii}{2586}). The identity of these possible blends remains unknown.

\subsection{Voigt profile fitting approach}\label{ss:fitting} 

We fitted the absorption profiles using the same general approach as employed in many previous works \citep[e.g.][]{Murphy_2003MNRAS.345..609M,King_2012MNRAS.422.3370K,Molaro_2013A&A...560A..61M}, particularly \citet{Kotus_2017MNRAS.464.3679K} who studied UVES spectra of the same absorber. We therefore present only a brief summary here, leaving further details and discussion about the fits for the three different regions in Sections \ref{ss:model_right}--\ref{ss:model_left}.

For each region, model profiles of the multi-component velocity structure of each transition were fitted to the ESPRESSO spectrum using {\sc vpfit} \citep[version 12.1;][]{Carswell_2014ascl.soft08015C}. This is a non-linear least-squares $\chi^2$ minimisation code designed specifically for fitting Voigt profiles to quasar absorption complexes. For a single ionic species, a velocity component is described by three free parameters -- its column density, Doppler $b$-parameter and redshift -- plus several atomic parameters of the transition determined from laboratory measurements or theory: wavelength, oscillator strength, damping constant and $Q$ coefficient. We use the atomic parameters compiled in \citet{Murphy_2014MNRAS.438..388M}, including the isotopic structures assuming terrestrial isotopic abundances (see further discussion in \Sref{sss:isotopes}). To measure \daa, the corresponding components of different ionic species must be assumed to share the same redshift -- see \Eref{e:alpha}. While this assumption has been tested in many previous works, we do so again with our higher resolving power ESPRESSO spectrum in \Sref{sss:shifts} below. The model profile is convolved with a Gaussian function representing the ESPRESSO instrumental line-shape. The Gaussian width is set to the mean resolution derived from fits to the LFC modes in the extracted wavelength calibration spectra: 2.10\,\kms\ for the blue arm and 2.05\,\kms\ for the red arm.

The velocity structure is not known \textsl{a priori}, so we take the same data-driven approach used in many recent studies: fitting as many velocity components as the data requires, as determined using $\chi^2$ per degree of freedom, $\chi^2_\nu$, as the information criterion \citep[e.g.][]{Evans_2014MNRAS.445..128E,Murphy_2016MNRAS.461.2461M,Murphy_2017MNRAS.471.4930M,Kotus_2017MNRAS.464.3679K}. That is, we attempt to fit many different models, with different numbers of components, and accept the one with the lowest $\chi^2_\nu$ as the fiducial model. In this process, we further assume that the $b$-parameter of a velocity component is the same for all ionic species. This assumption of turbulent broadening has been discussed extensively in the previous works cited above, but we discuss it further in \Sref{ss:structure} below. Finally, all $b$-parameters are limited to the range 0.15--30\,\kms\ during the $\chi^2$ minimisation process: the lower limit ensures components are not too much narrower than the $\approx$2\,\kms\ ESPRESSO resolution element, while the upper limit ensures that weak components do not become proxies for local continuum level changes in some parts of the fitting region.

After the fiducial model is established, only then is \daa\ introduced as a final free parameter (i.e.\ it is fixed at zero during the construction and comparison of models above). $\chi^2$ is minimised for that model again, providing the final best-fit parameter values, including \daa, plus their 1$\sigma$ uncertainties derived from the diagonal terms of the covariance matrix. Note that \daa\ is not strongly covariant with any other parameter because a single value applies to all components of all fitted transitions in a region: as $\alpha$ varies during the $\chi^2$ minimisation process, it shifts all components of a transition together, by the same velocity. This is fundamentally different to how the model responds to variation in the column densities, $b$-parameters or redshifts of individual components.

\subsection{Blind analysis}\label{ss:blinding}

Given the complexity of the $\zab=1.1508$ absorber's profile, it is
not possible with the above, human-directed model construction procedure to explore every potential fit. Decisions about where to place new components, to create more detailed fits, will be somewhat subjective. Of course, measuring \daa\ is our main goal, so we focussed more attention on fitting the narrowest features comprehensively, as these will most strongly constrain \daa; this reduces the number of possible fits worth exploring in detail. We then use an objective information criterion to select the fiducial model before \daa\ is measured. Nevertheless, it may seem plausible that some human biases remain. In principle -- though, we would argue, not in any practical way -- this may lead, or even be perceived to lead, to a biased measurement of \daa. To avoid any potential for this, we conducted the entire analysis in a blind manner such that it -- particularly the model construction step -- could not have been influenced by the results obtained.

The procedure for blinding the analysis followed that of several previous works \citep{Evans_2014MNRAS.445..128E,Murphy_2017MNRAS.471.4930M,Kotus_2017MNRAS.464.3679K}. Briefly, the ESPRESSO spectrum of HE\,0515$-$4414 was `blinded' in \textsc{uves\_popler} by introducing both long-range and intra-order distortions of the wavelength scale to each exposure. The size of the distortions varied randomly from exposure to exposure, but also contained some common features (also randomly assigned) to ensure the effects did not average away once the exposures were combined. Importantly, the size of these effects in the individual exposures and the final spectrum was not known, but their maximum possible size was limited so that they would lead to velocity shifts between the transitions corresponding to a spurious $\left|\daa\right| \la 5$\,ppm, about 4 times the final statistical uncertainty. This limit ensured that the model construction procedure was not unduly affected by the unknown shifts introduced between transitions.

All aspects of the analysis after the ESPRESSO DRS data reduction were conducted using the blinded spectra, i.e.\ \textsc{uves\_popler} processing, profile model construction, $\chi^2$ minimisation, parameter estimation (including \daa), systematic error analysis and consistency checks. Only once these steps were fully developed, refined and finalised was the blinding lifted in \textsc{uves\_popler} and the entire analysis re-run, without any further changes or human intervention, to produce the final results reported here.

\subsection{Model for the right region}\label{ss:model_right}

\begin{figure*}
\begin{center}
\includegraphics[width=0.98\textwidth]{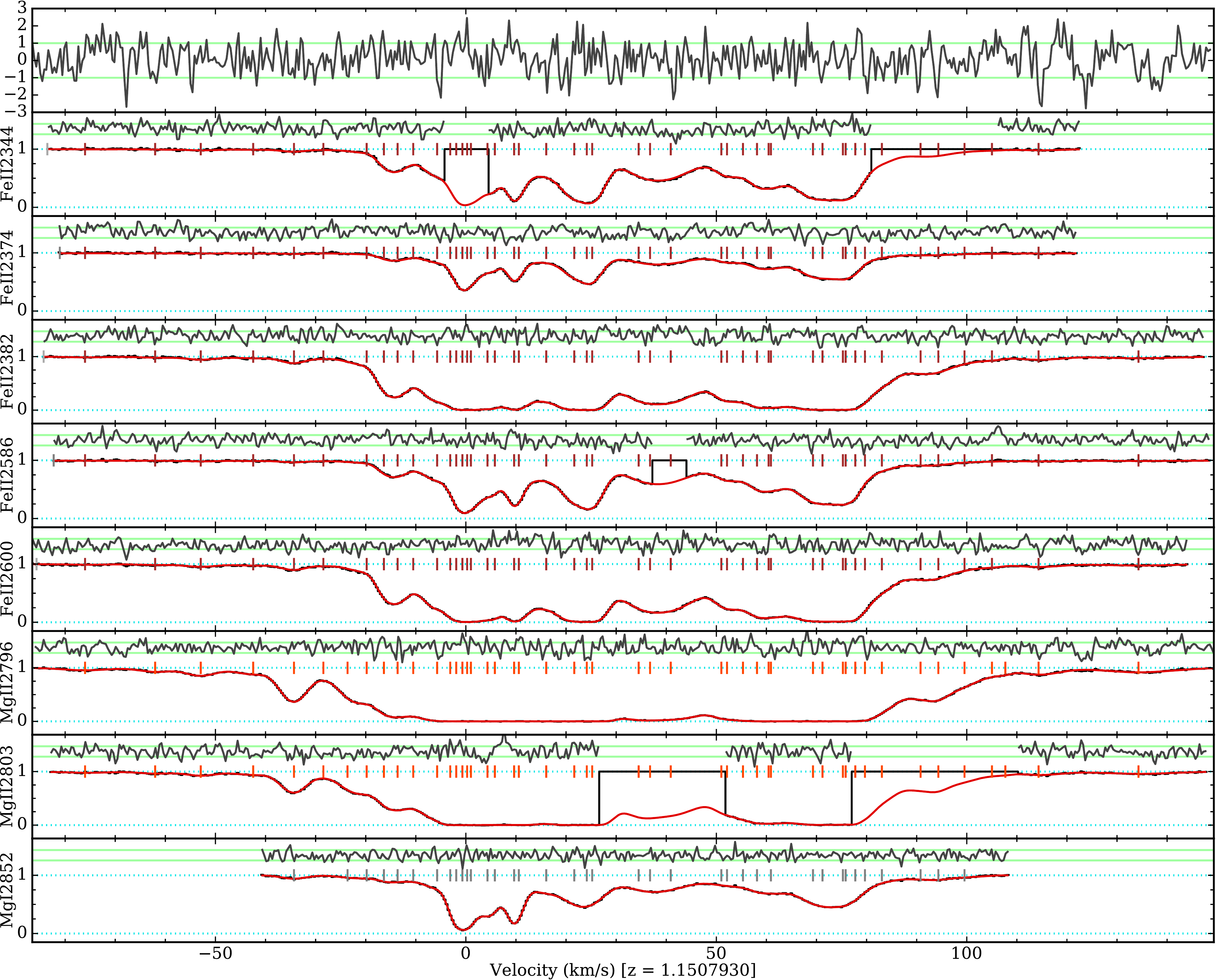} 
\vspace{-0.5em}
\caption{Combined, normalised ESPRESSO spectrum (black histogram) in the right region of the $\zab=1.1508$ absorber, for the strongest transitions only (the weak transitions are shown in \Fref{f:fit_r_weak}). Each panel shows a different transition, as labelled on the vertical axis. The red, solid line represents the fiducial model constructed from the Voigt profile components whose centroids are shown by tick-marks above each transition. Note that the fit traces the data so closely that the former almost obscures the latter in most places. To assist visualising any discrepancies, the normalised residuals [i.e.\ (data$-$model)/1$\sigma$ uncertainty] are shown above each transition (black line) with the $\pm1\sigma$ deviations marked by the green horizontal lines. The top panel shows the composite residual spectrum, i.e.\ the average normalised residual spectrum of the transitions shown, in units of standard deviations. The data missing from \tran{Mg}{ii}{2803}, \tran{Fe}{ii}{2586} and \tran{Fe}{ii}{2344} were excised manually due to the presence of additional absorption, though only the latter is confirmed to arise from a known absorber; see text for discussion.}
\label{f:fit_r_strong}
\end{center}
\end{figure*}

\begin{figure*}
\begin{center}
\includegraphics[width=0.98\textwidth]{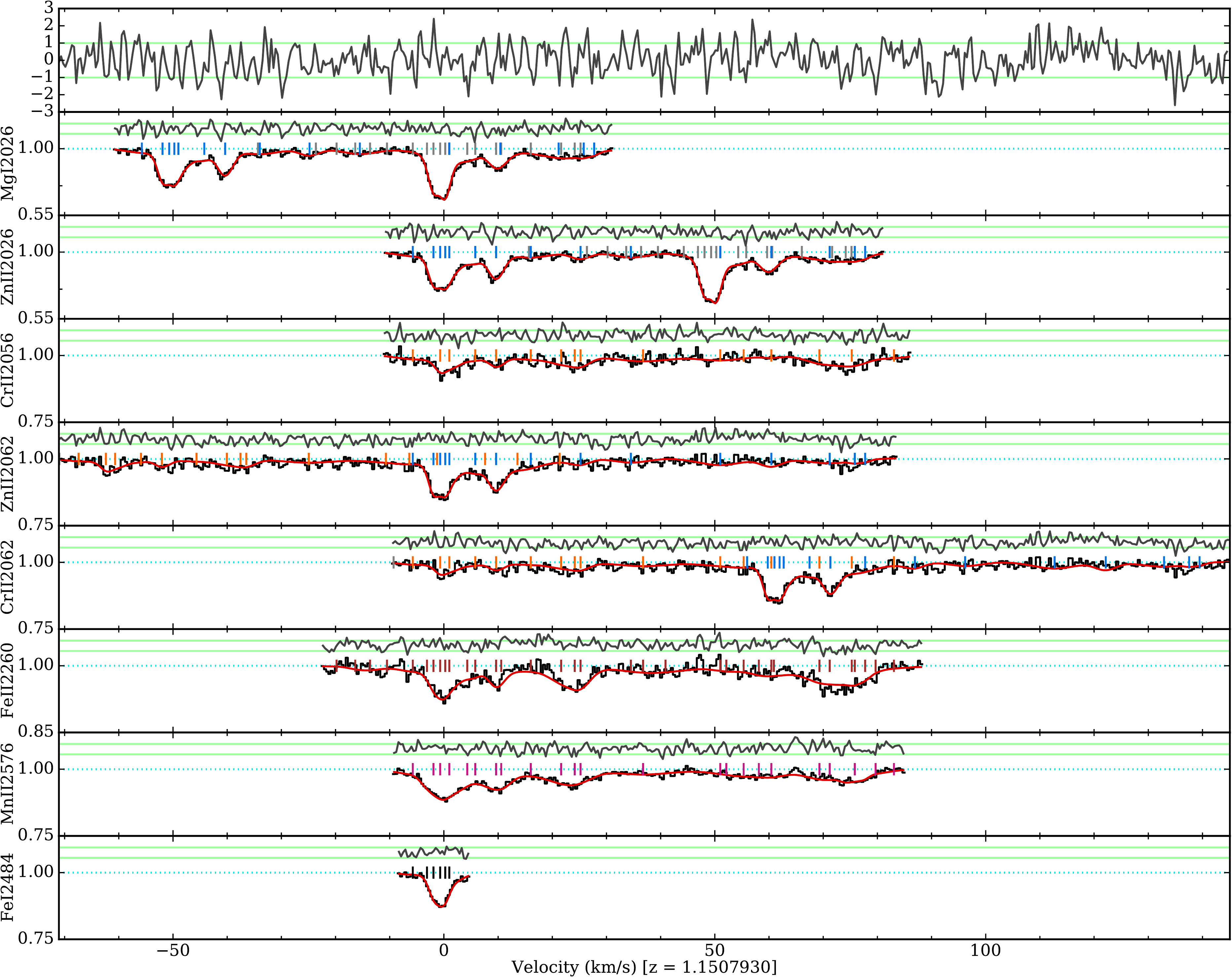} 
\vspace{-0.5em}
\caption{Same as \Fref{f:fit_r_strong} but for the weak transitions in the right region of the absorber. For simplicity, the weakest fitted transitions of \ion{Fe}{i}, \ion{Cr}{II} and \ion{Mn}{ii} are not plotted, but they were included in the fit: \tran{Fe}{i}{2523}, \tran{Cr}{ii}{2066}, \tran{Mn}{ii}{2594} and 2606.}
\label{f:fit_r_weak}
\end{center}
\end{figure*}

Figures \ref{f:fit_r_strong} and \ref{f:fit_r_weak} show the fiducial absorption profile model for the right region. The full {\sc vpfit} input and output files, providing all the parameter values and uncertainties, are provided in \citet{Murphy_2021Spec+Fits_HE0515} online. The strongest transitions, shown in \Fref{f:fit_r_strong}, have much higher \emph{effective} \SN\ -- i.e.\ the ratio of feature depth to \SN\ in the continuum -- and so provide much stronger constraints on \daa\ than the weak transitions in \Fref{f:fit_r_weak}. The main statistics of the fit are summarised in \Tref{t:fit_comp}, including the number of components fitted in each region. The model construction process was the most challenging for the right region. Indeed, the combination of high \SN, high resolving power, large number of transitions, and the narrow-but-clearly multi-component absorption features, made it the most difficult fit we have ever attempted. A total of 48 velocity components were required to fit the strongest transitions (the \ion{Mg}{ii} doublet) in the right region. \Tref{t:fit_comp} shows that this is similar to the number used by \citet{Kotus_2017MNRAS.464.3679K} to fit their UVES spectrum (49 components), but significantly more than used by \citet{Milakovic_2021MNRAS.500....1M} for their HARPS spectrum in the same region (26). This is unsurprising, to a first approximation, because the \SN\ per \kms\ for the ESPRESSO and UVES spectra are similar ($\approx$170 and 230, respectively) and much higher than for the HARPS spectrum ($\approx$58). However, we note that our fitting approach was the same as \citeauthor{Kotus_2017MNRAS.464.3679K}'s but quite different to \citeauthor{Milakovic_2021MNRAS.500....1M}'s; this is discussed further in \Sref{ss:structure}.

\begin{table*}
\caption{Final \daa\ measurements in the three regions, and statistics of the fits, compared with those of \citet{Kotus_2017MNRAS.464.3679K} and \citet{Milakovic_2021MNRAS.500....1M}. The best-fit values \daa\ and their statistical uncertainties are in units of parts-per-million (ppm). Note that the full results of this work, with systematic uncertainties included, are presented later in \Sref{s:results} (\Tref{t:results}). The \chisq\ per degree of freedom ($\chisq_\nu$) and number of fitted velocity components ($N_{\rm comp}$) in each model are also presented. The final row provides the weighted mean \daa\ and 1$\sigma$ uncertainty for our work. \citeauthor{Milakovic_2021MNRAS.500....1M} defined five regions, instead of the three used here and in \citeauthor{Kotus_2017MNRAS.464.3679K}; the values reported below for \citeauthor{Milakovic_2021MNRAS.500....1M} are representative only and combine their table 2's LFC values for regions I and II for our left region, and their regions III and IV for our central region, using the inverse of their total \daa\ uncertainties as weights.}\label{t:fit_comp}
\vspace{-1.5em}
\begin{center}
\begin{tabular}{l|cccc|cccc|cccc}
\hline\hline
        & \multicolumn{4}{c}{This work (ESPRESSO)} & \multicolumn{4}{c}{\citeauthor{Kotus_2017MNRAS.464.3679K} (UVES)} & \multicolumn{4}{c}{\citeauthor{Milakovic_2021MNRAS.500....1M} (HARPS)} \\
Region  & \daa  & $\sigma(\daa)$ & $\chisq_\nu$ & $N_{\rm comp}$ & \daa  & $\sigma(\daa)$ & $\chisq_\nu$ & $N_{\rm comp}$ & \daa  & $\sigma(\daa)$ & $\chisq_\nu$ & $N_{\rm comp}$ \\ 
\hline
Left    & 2.2   & 3.3            & 0.80        & 41                & -6.2   & 3.4           & 1.09         & 32                &  7.2  & 3.1            & 0.99         & 19               \\
Central & 1.6   & 5.6            & 0.84        & 40                & -4.7   & 1.7           & 1.17         & 25                & -4.7  & 4.0            & 0.87         & 17               \\
Right   & 1.1   & 1.5            & 1.03        & 48                & -0.8   & 0.6           & 1.26         & 49                & -2.6  & 3.4            & 0.99         & 26               \\
All     & 1.3   & 1.3            &             & 129               & -1.4   & 0.6           &              & 106               &  0.9  & 2.0            &              & 62               \\
\hline
\end{tabular}
\end{center}
\end{table*}

Despite the complexity and difficulty of the fitting process, the final \chisq\ of the fit, per degree of freedom, is very close to unity, $\chisq_\nu = 1.03$, indicating it is statistically acceptable. Importantly, we also do not observe any large outliers, or any remaining coherent structure in the residuals between the profile model and the data; these are normalised by the 1$\sigma$ flux uncertainty spectra and plotted above each transition in Figs.\ \ref{f:fit_r_strong} and \ref{f:fit_r_weak}. Finally, a `composite residual spectrum' is plotted above each figure, showing the average of the normalised residual spectra for the transitions plotted. These were calculated in the same way described in \citet{Malec_2010MNRAS.403.1541M} and offer additional power to detect evidence of statistically significant unfitted structure. However, we do not find any such evidence in the composite residual spectra for the right region.

Several gaps are apparent in the strong transitions of the right region in \Fref{f:fit_r_strong}. These were excised manually in {\sc uves\_popler} because of apparent blending. However, as noted in \Sref{ss:systems} above, only the blend in \tran{Fe}{ii}{2344} is associated with a known absorber (\tran{Ca}{ii}{3934} at $\zab=0.2818$). We detected very subtle, but highly statistically significant additional absorption in both \tran{Mg}{ii}{2803} and \tran{Fe}{ii}{2586} in the ESPRESSO spectrum. We confirmed its presence, over the same velocity ranges, in the UVES spectrum of \citet[][see their figure 4]{Kotus_2017MNRAS.464.3679K}, indicating that this is not telluric absorption. However, despite extensive searches for additional absorbers, and even considering possible diffuse interstellar band absorption in our Galaxy, we have not been able to identify these putative blends.

\begin{figure}
\begin{center}
\includegraphics[width=0.98\columnwidth]{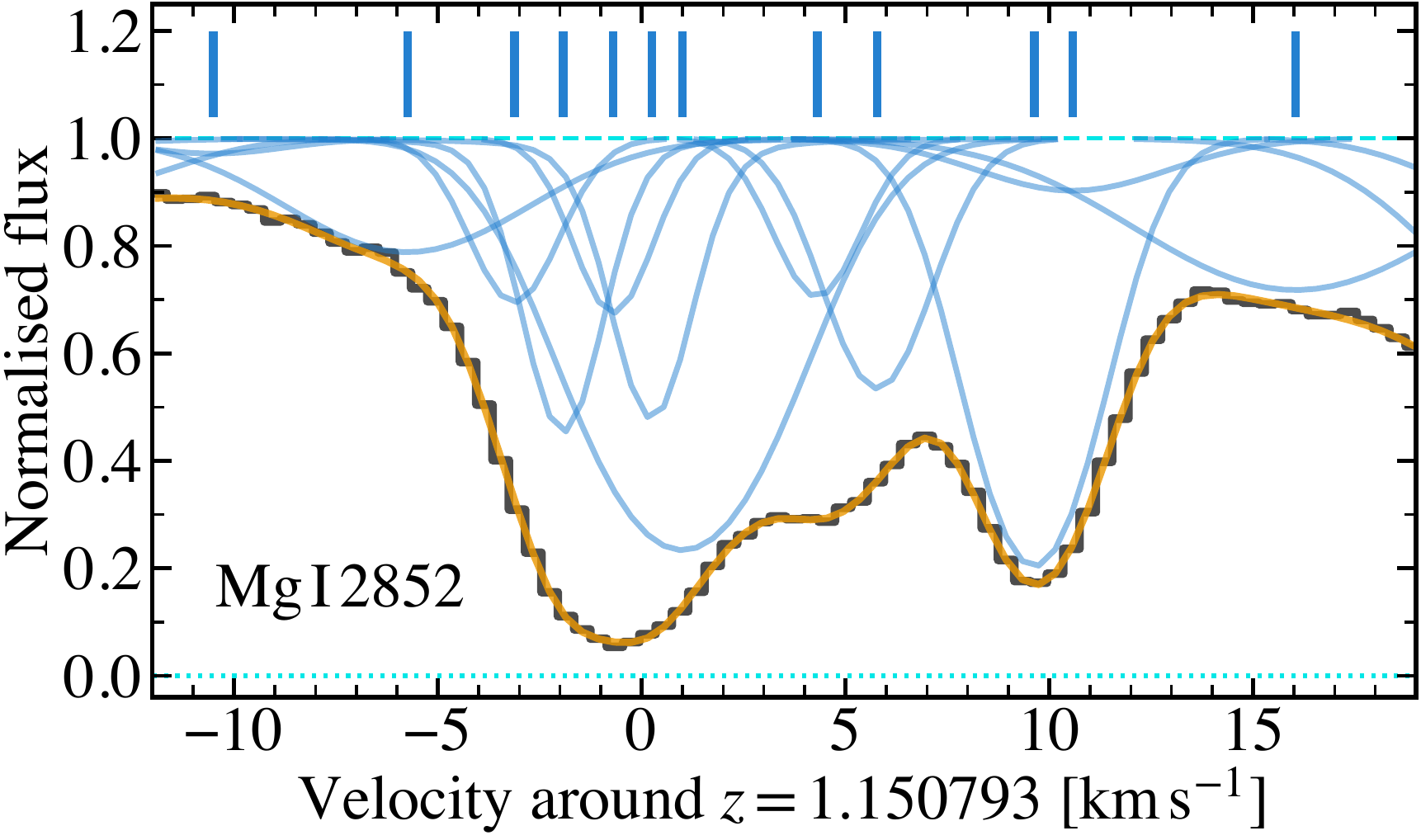}
\vspace{-0.5em}
\caption{Detailed \tran{Mg}{i}{2852} velocity structure showing the very narrow components near 0\,\kms\ in the right region. The model fit (solid orange line) overlays the spectrum (black histogram) and the Voigt profiles of individual components are shown in thin blue lines, with tick marks above the spectrum indicating their centroids. In our fiducial fit shown here, the components at $\approx-2$ and 0\,\kms\ have best-fit $b$-parameters of just $0.29 \pm 0.07$ and $0.22 \pm 0.13$\,\kms.}
\label{f:r_MgI_VPcomp}
\end{center}
\end{figure}

\begin{figure}
\begin{center}
\includegraphics[width=0.98\columnwidth]{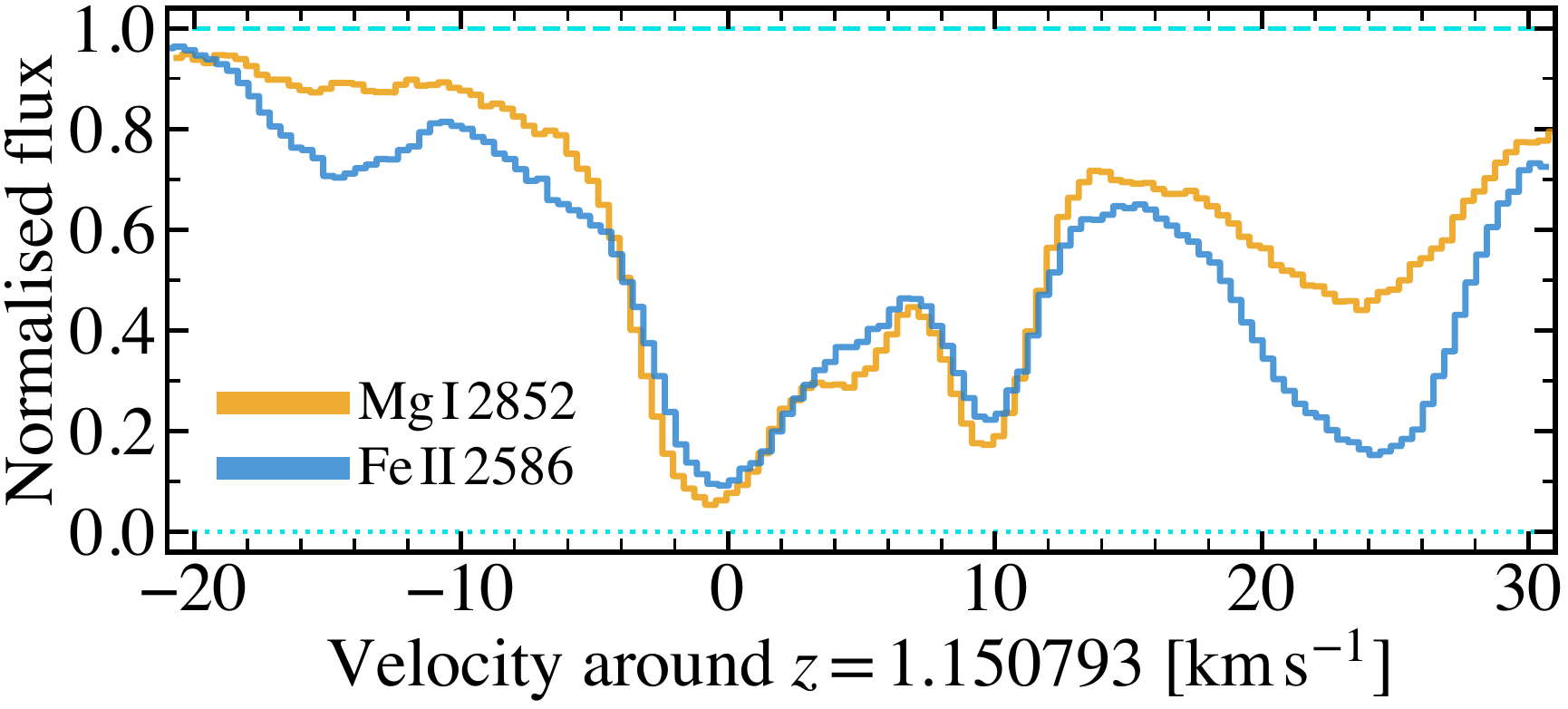}
\vspace{-0.5em}
\caption{Comparison between the \ion{Mg}{i} and \ion{Fe}{ii} velocity structures in the features at 0 and 9\,\kms\ in the right region. Note that the \ion{Mg}{i}/\ion{Fe}{ii} line depth ratio is larger for these features compared to those at other velocities. This, and possibly a small apparent shift between the 0\,\kms\ features in the two ions, strongly indicates the presence of cold, narrow components.}
\label{f:r_MgIvFeII}
\end{center}
\end{figure}

The strongest, narrowest features at 0 and 9\,\kms\ in \Fref{f:fit_r_strong} clearly should offer excellent constraints on \daa, but were also very difficult to fit. For the 0\,\kms\ feature, this is because it contains at least one, and most likely two very narrow components, with Doppler parameters $b\la0.5$\,\kms. \Fref{f:r_MgI_VPcomp} shows the detailed velocity structure of \tran{Mg}{i}{2852} in this feature, with the Voigt profile of each velocity component in our fiducial fit shown explicitly; the two components at $\approx-2$ and 0\,\kms are both very narrow. At first, it may seem unlikely that $b$-parameters narrower than 1/3 of the resolution element are required. Indeed, we found that a satisfactory fit of the \ion{Mg}{ii} and \ion{Fe}{ii} transitions, plus the strong \tran{Mg}{i}{2852} transition only, is possible without invoking these very narrow, or `cold', components. However, this fit grossly under-predicted the absorption observed in the 0\,\kms\ feature of the weaker \tran{Mg}{i}{2026} transition. The $b$-parameter of the strongest component in the 9\,\kms\ feature was similarly required to be quite narrow ($b\approx1.6$\,\kms) and some alternative fits required it to be even narrower ($b\approx0.7$\,\kms, though these fits were ultimately unstable). But this feature is very weak in \tran{Mg}{i}{2026}, correspondingly weakening its constraint on the $b$-parameter. The effect of the narrow components is also illustrated in \Fref{f:r_MgIvFeII}. Compared to the other absorption features, those at 0 and 9\,\kms\ are particularly strong in \ion{Mg}{i} relative to \ion{Fe}{ii}. The former also appear to be slightly shifted with respect to the corresponding \ion{Fe}{ii} feature. These are both manifestations of unresolved, cold velocity components blended with broader surrounding ones. We found no simpler velocity structure than that in \Fref{f:r_MgI_VPcomp} could fit these two features in all fitted transitions shown in Figs.\ \ref{f:fit_r_strong} and \ref{f:fit_r_weak}. This illustrates the substantial benefit of the high resolving power of ESPRESSO ($R \approx 145000$) in this case: \Fref{f:spec_comp} clearly shows how, for example, UVES's lower resolving power did not reveal the effects of these narrow components, even in the very high \SN\ UVES spectrum used by \citet[][$\SN \approx 250$ per 1.3\,\kms\ pixel]{Kotus_2017MNRAS.464.3679K}. These narrow components were also not required by the automatic fitting algorithm employed by \citet{Milakovic_2021MNRAS.500....1M}, most likely because the \tran{Mg}{i}{2026} transition was not covered by their HARPS spectra.

Corroborating the evidence for the cold components in the ESPRESSO spectrum is the detection of corresponding \ion{C}{i} and H$_2$ absorption. \citet{Quast_2002A&A...386..796Q} detected two \ion{C}{i} components at 0 and 9\,\kms, finding their widths to be $b = 2.0$ and 3.5\,\kms, respectively. Given the much lower resolving power of their UVES spectra ($R \approx 55000$, i.e.\ 5.5\,\kms\ full-width-half-maximum resolution), it is possible, even likely these $b$-parameters are overestimated. From the column densities in the different \ion{C}{i} transitions, \citet{Quast_2002A&A...386..796Q} inferred  the populations of the fine-structure levels, finding the kinetic temperature to be $\sim$240\,K, i.e.\ thermal broadening of $b_{\rm ther} \approx 0.6$\,\kms. \citet{Reimers_2003A&A...410..785R} discovered molecular hydrogen absorption covering the redshift range of these two cold \ion{C}{i} components with Hubble Space Telescope spectroscopy ($R \sim 30000$), finding a total H$_2$ column density of $N({\rm H}_2) = 8.7^{+8.7}_{-4.0}\times10^{16}{\rm \,cm}^{-2}$. They also estimated the photodissociation rate to be more than 300 times the mean Galactic disk value, suggesting enhanced star-formation activity and a gas density of $\sim$100\,cm$^{-2}$. This high density implies a very small cloud size, $L \sim 0.25$\,pc, along the line-of-sight for the molecular cloud. The relative column densities of the first two rotational levels ($J=0$ and 1) of the ground electronic state provided an excitation temperature estimate of $T_{01} \sim 90$\,K. It is therefore unsurprising to find very narrow neutral and singly-ionised velocity components in our ESPRESSO spectrum, associated with the cold, small \ion{C}{i} and H$_2$ gas clouds, which are likely embedded in warmer, more rarefied gas which gives rise to other, broader components at very similar velocities. Interestingly, complexes of clouds with precisely these properties have been observed in our Galaxy through $R\approx10^6$ optical spectroscopy of interstellar \ion{Na}{i} and \ion{Ca}{ii} absorption lines \citep{Barlow_1995MNRAS.272..333B}.

The requirement to fit \tran{Mg}{i}{2026} to strongly constrain the narrow components' parameters implies the need to fit the five weak transitions of Zn and \ion{Cr}{ii}. Firstly, it is important to fit the \tran{Zn}{ii}{2026} transition which partially blends with \tran{Mg}{i}{2026} (the two transitions are separated by 50\,\kms). This can be seen in the top two spectra of \Fref{f:fit_r_weak}. Given this blending, constraining the \ion{Zn}{ii} parameters is greatly improved by fitting the \tran{Zn}{ii}{2062} transition as well, but this is blended with \tran{Cr}{ii}{2062} -- see \Fref{f:fit_r_weak}'s middle two spectra. To more reliably constrain \ion{Cr}{ii}'s parameters, we therefore need to fit \tran{Cr}{ii}{2056} and 2066 as well.

The remaining weak transitions fitted in the right region are the \ion{Mn}{ii} triplet and the two strongest transitions of \ion{Fe}{i}: $\lambda$2484 and 2523. To our knowledge, this is the first time \ion{Fe}{i} transitions have been included in a \daa\ measurement -- they are rarely even detected in quasar absorbers. While they are very weak in our ESPRESSO spectrum, their strongest features appear just as narrow as in most other transitions so we include them in our fit mainly to help constrain the $b$-parameters of the cold components discussed above. It is important to note that their laboratory wavelengths are only known with $\sim$38--74\,\ms\ precision \citep{Nave_1994ApJS...94..221N,Nave_2011JOSAB..28..737N}, which is another reason they are not normally included in \daa\ measurements. Nevertheless, they are only measured with effective velocity precisions of $\sim$300\,\ms\ each in our ESPRESSO spectrum -- see \Sref{sss:shifts}. This is much larger than the best constrained transitions in the right region ($\sim$30\,\ms), and their laboratory wavelength precisions, so including them will not cause significant systematic errors in \daa. We fitted the \ion{Mn}{ii} triplet for completeness, because it was detected, but in practice it provides especially weak constraints on \daa: the absorption in all three transitions is very weak and their broad hyperfine structure \citep[$\ga$4\,\kms\ wide;][]{Blackwell-Whitehead_2005MNRAS.364..705B,Aldenius_2009PhST..134a4008A} effectively smooths out the sharper features.

The strong saturation in the \ion{Mg}{ii} doublet meant that the column densities of several \ion{Mg}{ii} velocity components were unconstrained. While the $b$ parameters and redshifts of these components are, by assumption, directly linked to those of the other fitted ions, the column densities of the nine \ion{Mg}{ii} components between $-6$ and 18\,\kms, and the three between 133 and 142\,\kms, had to be fixed to reasonable values to ensure they did not destabilise the fit (e.g.\ become very high during the $\chisq$ minimisation, leading to other components being rejected by \textsc{vpfit}). We estimated the fixed values by assuming the ratios of their column densities with nearby, unsaturated components matched the corresponding ratios in \ion{Fe}{ii}. This has a negligible impact on our results because the saturated regions do not directly constrain \daa. Almost identical results were obtained by artificially increasing the flux errors on the affected pixels, or by removing those pixels from the $\chisq$ minimisation process altogether. Similarly, fixing \daa\ to zero for the saturated components, or allowing it to take a different value than for the other, unsaturated components, yielded negligible changes to the results.

Finally, we note that $\chisq_\nu$ for the right region (1.0) is slightly larger than for the central and left regions (0.8) -- see \Tref{t:fit_comp}. At first, it may seem that this implies that our models of the central and left regions contain too many components, and are over-fitting the data, especially since $\chisq_\nu$ in the right region is so close to the expected value of unity. However, in our combined ESPRESSO spectrum the root-mean-square flux deviations around unabsorbed sections of continuum are, in general, $\approx$10\% smaller than the average formal statistical flux uncertainty in the same sections. At present, this slight overestimation of the flux error spectrum appears to arise due to a combination of effects in the DRS and {\sc uves\_popler} (e.g.\ rebinning leading to small correlations between the fluxes and flux uncertainties in neighbouring pixels). The \chisq\ values in the left and central regions are, therefore, expected in this case, while that in the right region may, instead, actually indicate a small degree of \emph{under}-fitting.

\subsection{Model for the central region}\label{ss:model_central}

\begin{figure*}
\begin{center}
\includegraphics[width=0.98\textwidth]{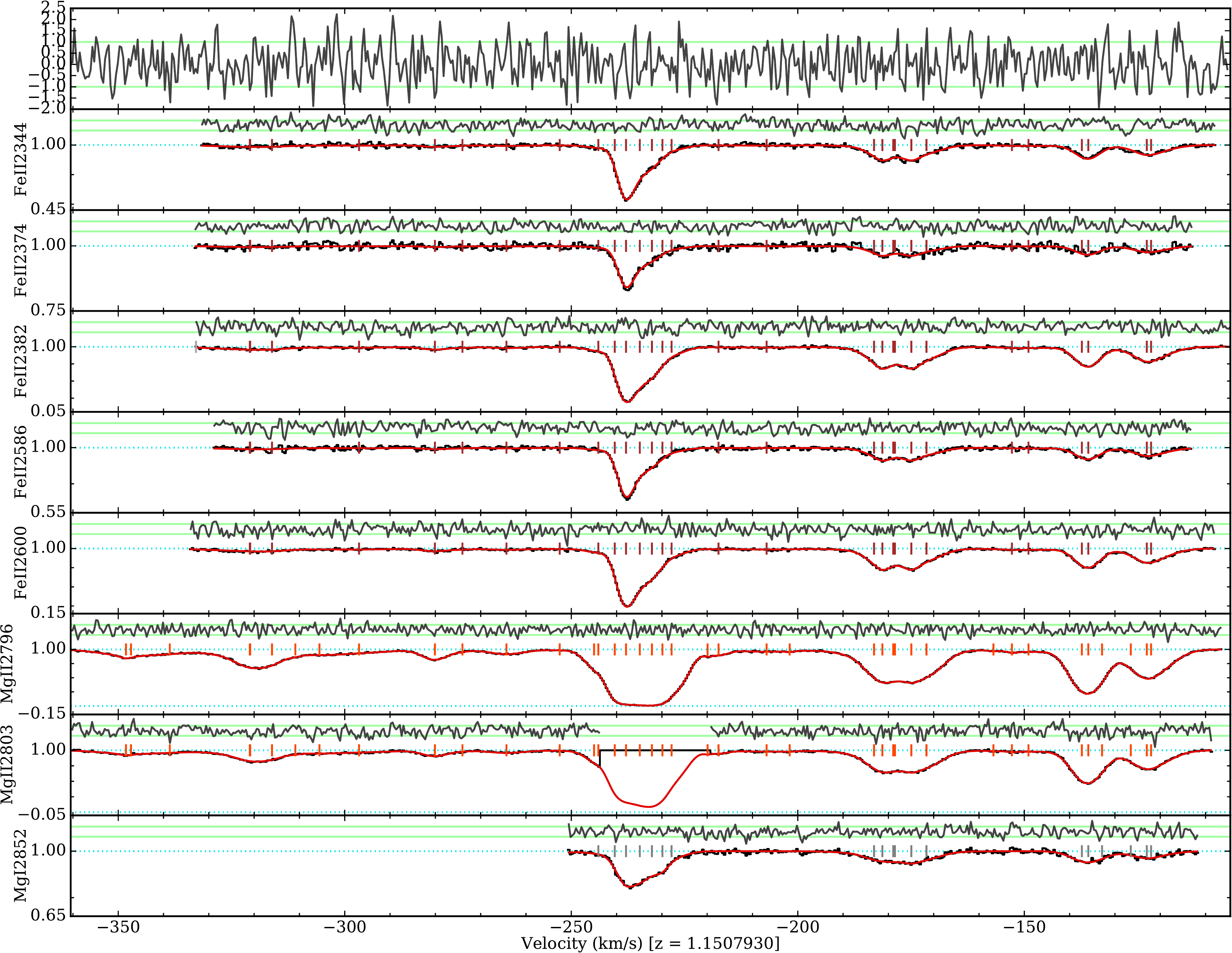} 
\vspace{-0.5em}
\caption{Same as \Fref{f:fit_r_strong} but for the central region of the absorber and for all transitions fitted. For clarity, the very weak (formally undetected) \tran{Mg}{i}{2026} line is not plotted but it was included in the fit -- see text for discussion.}
\label{f:fit_c}
\end{center}
\end{figure*}

\Fref{f:fit_c} shows our fiducial profile fit to the absorption lines in the central region. The absorption is substantially weaker than in the right region for all ions, and the weak transitions used in the right region are not detected here. The fiducial velocity structure was, again, very difficult to construct, requiring many alternatives to be explored. In this case, the \ion{Fe}{ii}/\ion{Mg}{ii} column density ratio varies strongly across the feature at $-237$\,\kms, making it more difficult to find a consistent velocity structure to fit to all ions. However, this was considerably simplified by the discovery that the two \ion{Mg}{ii} transition profiles were inconsistent with each other due to additional absorption near $-237$\,\kms\ in \tran{Mg}{ii}{2803}. It appears likely that this is associated with the additional absorption identified in the right region of the same transition but, similarly, we have not been able to identify its origin (see \Sref{ss:model_right} above).

Four main features provide the constraints on \daa\ in the central region: the strongest feature at $-237$\,\kms, the double-peaked feature at $-178$\,\kms, and the two moderately separated features at $-135$ and $-123$\,\kms. For the former, the saturation in the \ion{Mg}{ii} lines means that the \ion{Mg}{i} and \ion{Fe}{ii} transitions most strongly constrain \daa. For the other features, the \ion{Mg}{i} absorption is too weak so the \daa\ constraints come from the \ion{Mg}{ii} and \ion{Fe}{ii} transitions. The strongest absorption at $-237$\,\kms\ also appears to be the narrowest, but we note that very narrow, `cold' components were not required to fit this feature, in contrast to the 0 and 9\,\kms\ features in the right region. While not formally detected, we do include the \tran{Mg}{i}{2026} transition (not shown on \Fref{f:fit_c}) in the fit because it helps to rule out the presence of any very narrow components in this feature; removing it from the fit has a small effect on the $b$ parameter constraints for the velocity component fitted here, but a negligible effect on \daa, as expected.

\Tref{t:fit_comp} shows that the 40 components fitted to the central region provided a $\chisq_\nu = 0.84$, indicating a statistically satisfactory model. There is no strong evidence in the normalised residual spectra, plotted above the individual transitions in \Fref{f:fit_c}, of unfitted velocity structure. The composite residual spectrum (top panel) also shows no evidence of any remaining structure. In contrast to the right region, \Tref{t:fit_comp} shows that our fit to the ESPRESSO data required substantially more components than \citet{Kotus_2017MNRAS.464.3679K} used for the same region of their UVES spectrum (25 components). However, we note that their $\chisq_\nu$ was somewhat larger than ours, so it is possible that further exploration of alternative, more complex fits to the UVES spectrum would have yielded a statistically better fit. As in the right region, the fact that \citet{Milakovic_2021MNRAS.500....1M} fitted 17 components is likely due to the substantially lower \SN\ per \kms\ of their HARPS spectrum.

\subsection{Model for the left region}\label{ss:model_left}

\begin{figure*}
\begin{center}
\includegraphics[width=0.98\textwidth]{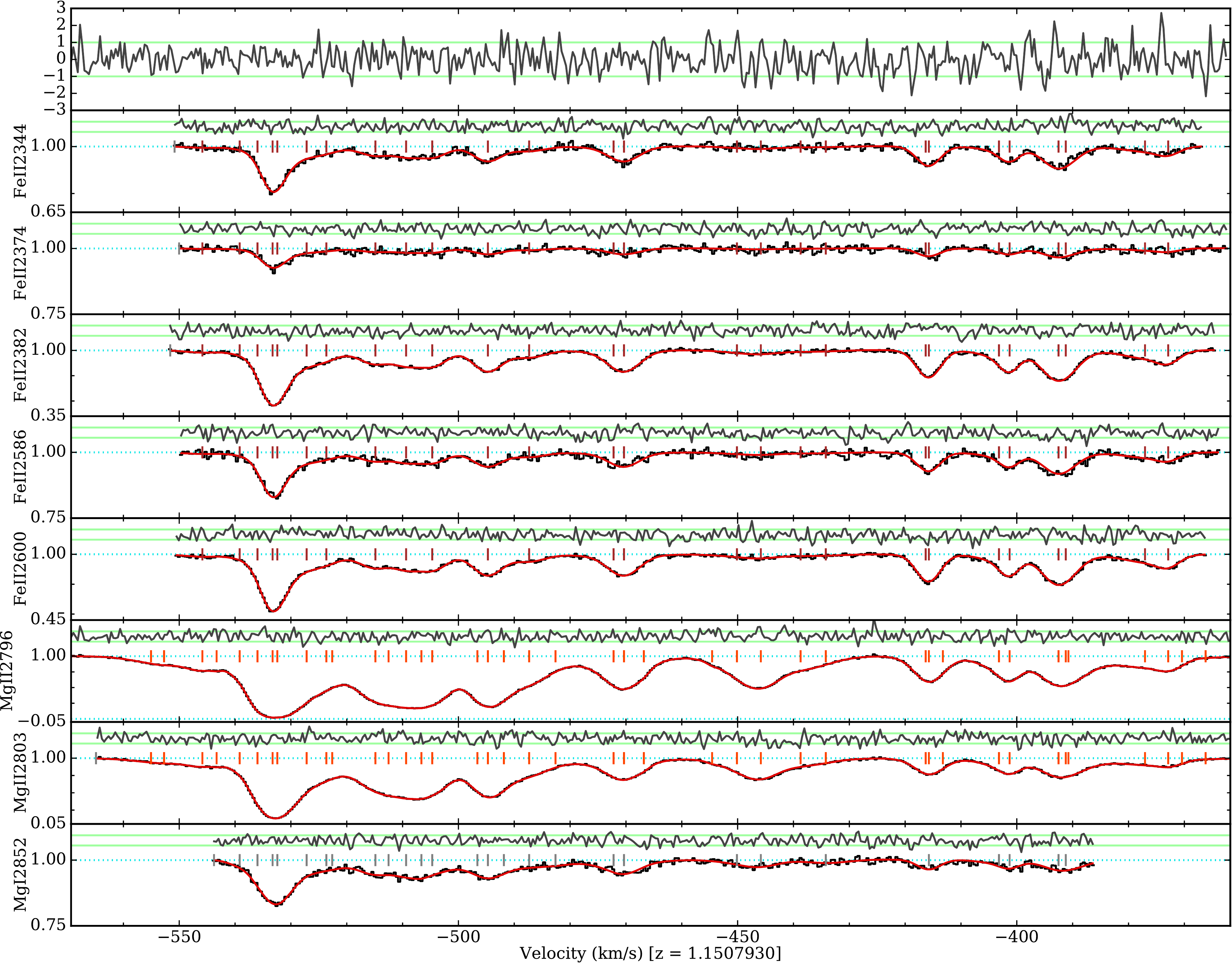} 
\vspace{-0.5em}
\caption{Same as \Fref{f:fit_r_strong} but for the left region of the absorber and for all transitions fitted.}
\label{f:fit_l}
\end{center}
\end{figure*}

\Fref{f:fit_l} shows our profile fit to the left region of the absorption system. Only the strongest transitions of \ion{Mg}{i}/{\sc ii} and \ion{Fe}{ii} are detected and used to constrain \daa\ here. While still challenging, this fit was considerably simpler to construct than those in the right and central regions. Nevertheless, the two features between $-520$ and $-480$\,\kms\ presented some difficulties, with the five strongest components seen in \ion{Fe}{ii} not being adequate to fit the \ion{Mg}{i}/{\sc ii} profiles. The latter required five additional, very weak components which appeared too weak in \ion{Fe}{ii} to be statistically required -- {\sc vpfit} rejected these from the \ion{Fe}{ii} velocity structure. Given that this occurs over a small-but-contiguous portion of the profile, it is likely that this represents a real effect, i.e.\ the \ion{Fe}{ii}/\ion{Mg}{ii} column density ratio may simply be lower in this part of the absorber. Indeed, visual comparison of the \ion{Fe}{ii}/\ion{Mg}{ii} absorption strength here with that in the strongest part of the region nearby, at $-532$\,\kms, supports this conclusion.

In this left region of the absorber there are numerous distinct features which all contribute constraints on \daa, but the combination of \ion{Mg}{i} and \ion{Fe}{ii} transitions in the strongest feature at $-582$\,\kms\ are the most constraining. The \ion{Mg}{ii} lines are somewhat saturated in that feature, and the surrounding components are, therefore, deeper; this effectively broadens the feature in \ion{Mg}{ii}, reducing its impact on \daa\ here. Nevertheless, \ion{Mg}{i} is much weaker in the other features across the profile, so the dominant \daa\ constraints arise from the combination of \ion{Mg}{ii} with \ion{Fe}{ii} there.

Our fit to the ESPRESSO spectrum required 41 components in the left region, with a final $\chisq_\nu = 0.80$, similar to the central region. As with the other two regions, there are no strong outliers revealed by the normalised residual spectra in \Fref{f:fit_l} and the composite residual spectrum shows no evidence for remaining, unfitted velocity structure. \Tref{t:fit_comp} again shows that we fitted substantially more velocity components than \citet{Kotus_2017MNRAS.464.3679K} and, particularly, \citet{Milakovic_2021MNRAS.500....1M}, but we consider the reasons discussed for the same situation in the central region (\Sref{ss:model_central} above) to apply here as well.

\section{Results}\label{s:results}

\subsection{Results with statistical uncertainties}\label{ss:results}

\Tref{t:fit_comp} provides the best-fitting values of \daa\ and 1$\sigma$ statistical uncertainties for the three regions of the $z=1.1508$ absorber. These are the results from {\sc vpfit} after minimising $\chisq$ between the fiducial models (described in Sections \ref{ss:model_right}--\ref{ss:model_left}) and the combined, unblinded ESPRESSO spectrum, as described in \Sref{ss:fitting}. Only the statistical uncertainties are provided here, as output by {\sc vpfit}, derived from the final parameter covariance matrix, i.e.\ for a given model, they are determined by the photon statistical noise in the fitted regions of the spectrum. The systematic error budget is assessed in detail below (\Sref{ss:syserr}).

The most important aspect of the results in \Tref{t:fit_comp} is that \daa\ is within the 1$\sigma$ statistical uncertainty of zero for all three regions. As expected, the right region provides the tightest constraint, with a statistical uncertainty of just 1.5\,ppm. The weighted mean from our ESPRESSO spectrum is also within its 1$\sigma$ uncertainty of zero, i.e.\ 1.3\,ppm. This overall statistical precision is very similar to the ensemble precision obtained from the large samples of absorbers from Keck \citep[1.1\,ppm; ][]{Murphy_2003MNRAS.345..609M,Murphy_2004LNP...648..131M} and VLT \citep[1.2\,ppm; ][]{Webb_2011PhRvL.107s1101W,King_2012MNRAS.422.3370K}. As explained in \Sref{s:intro}, the high likelihood of large systematic effects in those studies undermines confidence in the results. The more recent 26 measurements from Keck, VLT and Subaru which are corrected for, or insensitive to, these systematic effects also have a similar ensemble precision to our new ESPRESSO measurement \citep{Evans_2014MNRAS.445..128E,Murphy_2016MNRAS.461.2461M,Murphy_2017MNRAS.471.4930M}.

Nevertheless, \Tref{t:fit_comp} shows that the 1$\sigma$ uncertainty reported by \citet{Kotus_2017MNRAS.464.3679K} from their UVES spectrum of HE\,0515$-$4414 is approximately half of ours, despite the \SN\ per \kms\ being similar to that in our ESPRESSO spectrum. This is expected because the ESPRESSO spectrum does not cover the \tran{Fe}{ii}{1608} transition which provides strong-but-unsaturated absorption in the right region and, most importantly, a large \emph{negative} $Q$ coefficient. This transition, combined with the other strong \ion{Fe}{ii} transitions which have large \emph{positive} coefficients (see \Fref{f:Qvswl}), provided the strongest constraint on \daa\ in the UVES spectrum. Indeed, \citet{Kotus_2017MNRAS.464.3679K} predicted that an ESPRESSO spectrum with similar \SN\ per \kms\ as the UVES spectrum would provide a statistical precision of $\sim$1.8\,ppm.
%We reassess the statistical precision derived from the UVES spectrum in \Sref{ss:previous} given the presence of the `cold' components identified in our profile fit to the ESPRESSO data.

\subsection{Systematic  errors}\label{ss:syserr}

While the {\sc vpfit} \chisq\ minimisation process provides a robust statistical uncertainty for a given model fit, it is clearly important to consider and quantify any systematic effects at the $\le$1\,ppm level. These have been considered in detail in many previous works \citep[e.g.][]{Murphy_2001MNRAS.327.1223M,Murphy_2003MNRAS.345..609M,Fenner_2005MNRAS.358..468F,Levshakov_2005A&A...434..827L,Levshakov_2007A&A...466.1077L,Molaro_2008EPJST.163..173M,Griest_2010ApJ...708..158G,Whitmore_2010ApJ...723...89W,Rahmani_2013MNRAS.435..861R,Evans_2014MNRAS.445..128E,Whitmore_2015MNRAS.447..446W}. In the slit-based echelle spectrographs on the Keck, VLT and Subaru telescopes, light-path differences between the quasar and ThAr calibration light are most likely responsible for the largest systematic effects: both long-range and intra-order distortions of the wavelength scale. These can produce spurious shifts in \daa\ up to $\sim$10\,ppm in typical absorbers \citep{Evans_2014MNRAS.445..128E}. When measuring \daa\ from a single absorber, as we do here, it is vital that systematic errors are suppressed and any remaining ones assessed. In their UVES study of HE0515$-$0515, \citet{Kotus_2017MNRAS.464.3679K} used the existing HARPS spectrum of the same quasar to correct for the long-range distortions of UVES's wavelength scale. Nevertheless, uncertainties in this recalibration, and the intra-order distortions, meant that wavelength calibration errors remained the dominant factor in their 0.65\,ppm systematic error budget. As discussed in \Sref{s:intro}, the LFC calibration of ESPRESSO effectively removes this dominant source of systematic uncertainties in \daa\ measurements.

The remaining systematic effects in our measurements arise from three aspects of the data processing and analysis: spectral redispersion, profile modelling and convergence of the \chisq\ minimisation process. These have been discussed in several previous works, including for HE\,0515$-$4414 in \citet{Kotus_2017MNRAS.464.3679K}, and we discuss and quantify each of them below using the same approaches as that study. The results are summarised in \Tref{t:results}.

\begin{table}
\caption{Summary of \daa\ values from this work for the left, central and right regions, in units of ppm. Their 1$\sigma$ statistical and systematic error contributions are described in \Sref{ss:results} and \Sref{ss:syserr}, respectively. The combined value of \daa\ is the weighted mean from the three regions, with 1$\sigma$ statistical and systematic error values, as described in \Sref{ss:comb}: $\daa = 1.31 \pm 1.29_{\rm stat} \pm 0.43_{\rm sys}$\,ppm. Note that all quantities are quoted with extra significant figures only to allow reproducibility. The checks on additional systematic errors in \Sref{ss:consistency} were conducted in the right region only, with no evidence found that they contribute to the systematic error budget. Section numbers are provided in parentheses for the systematic error checks.}
\label{t:results}
\vspace{-1.5em}
\begin{center}
\begin{tabular}{lcccc}
\hline\hline
                                      & \multicolumn{3}{c}{Region}                         &                 \\
Result type                           & Left        & Central               & Right        &  Comb.          \\
\hline
\multicolumn{5}{l}{\textbf{Fiducial fitting results}}                                                        \\
$\daa$ value                          & 2.17        & 1.57                  & 1.14         &  \textbf{1.31}  \\
1$\sigma$ statistical error           & 3.31        & 5.59                  & 1.45         &  \textbf{1.29}  \\
\multicolumn{5}{l}{\textbf{Systematic errors}}                                                               \\
Redispersion (\ref{sss:redisp})       & 0.62        & 0.10                  & 0.24         &  0.21           \\
Profile modelling (\ref{sss:profile}) & 1.19        & 2.26                  & 0.23         &  0.28           \\
Convergence (\ref{sss:convergence})   & 0.18        & 0.72                  & 0.31         &  0.25           \\
Combined                              & 1.35        & 2.37                  & 0.45         &  \textbf{0.43}  \\
\multicolumn{5}{l}{\textbf{Systematic error checks}}                                                         \\
\multicolumn{3}{l}{LFC vs.\ ThFP calibration (\ref{sss:LFCvsThFP})}         & $<$0.02      &                 \\
\multicolumn{3}{l}{Combined LFC$+$ThFP calibration (\ref{sss:combLFCThFP})} & $<$0.01      &                 \\
\multicolumn{3}{l}{Arm shifts (\ref{sss:arms})}                             & $<$0.14      &                 \\
\multicolumn{3}{l}{Trace shifts (\ref{sss:traces})}                         & $<$0.16      &                 \\
\multicolumn{3}{l}{Isotopic abundance variations (\ref{sss:isotopes})}      & $<$0.7       &                 \\
\hline
\end{tabular}
\end{center}
\end{table}

\subsubsection{Spectral redispersion effects}\label{sss:redisp}
 
To combine the extracted spectra from multiple exposures, {\sc uves\_popler} defines a final (log-linear) wavelength grid onto which it redisperses the flux from each trace in each exposure. The phases of the initial and final bins are different, so the fluxes (and uncertainties) in neighbouring pixels are somewhat correlated in the final spectrum. This can lead to small errors in the best-fit centroid position of a spectral feature and, therefore, systematic errors in \daa. To quantify this effect, we created 10 alternative realisations of the combined ESPRESSO spectrum with {\sc uves\_popler} with dispersions between 0.395 and 0.405\,\kms\ per pixel (i.e.\ differing by 1\,\ms\ per pixel) and re-computed \daa. For each realisation, the fiducial model was re-optimised with \daa\ initially fixed to the fiducial result (from the 0.4\,\kms\ per pixel spectrum), then run again with \daa\ as a free parameter.

The resulting redispersion uncertainty for each region is provided in \Tref{t:results}. The systematic uncertainty is taken as the standard deviation in the difference between \daa\ for the realisations and the fiducial value for the region. As expected, \Tref{t:results} shows that the redispersion errors are small compared to the statistical uncertainties for all regions; for the right region, which is most important for determining \daa, the redispersion effect is $\la$20\% of the statistical uncertainty. It is possible that the redispersion error is underestimated in the central region (0.1\,ppm) because the convergence error was larger for that model -- see \Sref{sss:convergence} below.

\subsubsection{Profile modelling uncertainties}\label{sss:profile}

While our fiducial model provides a statistically acceptable fit, we cannot be sure that our fitted velocity structure is unique. This model uniqueness problem is discussed further in \Sref{ss:structure} after we compare our result with others in the literature (\Sref{ss:previous}). However, as described in \Sref{ss:fitting} above, we explored an extensive range of alternative models while trying to determine which one provided the lowest \chisq\ per degree of freedom for each region. Similar to the approach of \citet{Kotus_2017MNRAS.464.3679K}, we can use this set of alternative fits to estimate the systematic uncertainty in our fiducial \daa\ measurements from model non-uniqueness.

For each region, there is an enormous number of possible alternative models, but the vast majority of these will provide much poorer fits to the data than our fiducial one. Therefore, we considered alternative models that contained only one more or fewer velocity component than our fiducial model. We also focussed most attention on the spectral features that most strongly constrained \daa\ in each region (see Sections \ref{ss:model_right}--\ref{ss:model_right}). For example, the strong, narrow features near 0\,\kms\ in the right region were thoroughly explored, while the weak, broad feature at 30--50\,\kms\ is much less important. \Aref{s:models} provides the basic properties of the alternative models in \Tref{t:models}. Given our fitting approach (\Sref{ss:fitting}), many of these models were not acceptable: they produced large, correlated structures in the composite residual spectra, or were not `stable', i.e.\ {\sc vpfit} rejected components in one species that were relatively strong in other species. We did not consider these alternative models any further. Finally, we considered the remaining alternative fits that had an Akaike information criterion, corrected for finite sample sizes, AICC \citep[][as implemented in \citealt{King_2012MNRAS.422.3370K}]{Akaike_1974ITAC...19..716A,Sugiura_1978} no more than 10 higher than the fiducial model's AICC. According to the Jeffreys' scale for interpreting the AICC, a difference of 10 corresponds to `very strong' evidence against the model with the higher AICC \citep{Jeffreys_1961}. This final selection criterion yielded a sample of 5, 6 and 6 alternative models for the right, central and left regions, respectively.

The profile modelling error derived from the alternative models is shown for each region in \Tref{t:results}. The systematic uncertainty is taken as the standard deviation in the difference between \daa\ for the alternative models and the fiducial value for the region. Remarkably, we find the modelling uncertainty is smaller than the statistical uncertainty in all regions, especially in the right region where the fiducial fit was the most challenging to construct -- there it is only $\la$20\% of the statistical uncertainty. Two additional alternative models for the right region had AICC differences of 11 and 12 -- only slightly larger than our accepted threshold of 10 -- but relatively small deviations in \daa\ from the fiducial value; including them in the results would not significantly alter the systematic uncertainty estimate. We discuss this consistency between alternative models in light of recent literature in \Sref{ss:structure} below. It is important to emphasise here that all the alternative fits were constructed based on the blinded spectrum, with only the AICC criterion discussed above applied using the final, unblinded \chisq\ and AICC values, thereby removing any possible (though highly unlikely) human biases in estimating this systematic uncertainty.

\subsubsection{Convergence errors}\label{sss:convergence}

As can be seen in Figs.\ \ref{f:fit_r_strong}--\ref{f:fit_l}, a very large number of velocity components were required to fit the high \SN\ ESPRESSO spectrum, many of which are within a resolution element of neighbouring components. This creates degeneracies between their parameters (i.e.\ their column densities, $b$-parameters and redshifts) in the \chisq\ minimisation process: it means \chisq\ will have a weak dependence on these parameters, causing the minimisation algorithm to take smaller steps in all model parameters. This can mean the \chisq\ difference between iterations can fall below the user-supplied stopping criterion before all parameters have completely reached their true, best-fit values. We measure the uncertainty this convergence error causes by re-measuring \daa\ after offsetting its initial guess from the fiducial value in \Tref{t:results} by $\approx$1$\sigma_{\rm stat}$. More specifically, we first re-optimised the fit for a given region with \daa\ fixed to its offset starting value, and then re-ran the \chisq\ minimisation with \daa\ as a free parameter.
 
\Tref{t:results} shows the convergence uncertainty for each region, which was taken as half the difference between the \daa\ values measured after offsetting the starting point higher and lower than the fiducial value. For each region, the convergence uncertainty is small compared to the statistical uncertainty for that region. For the most important right region, \daa\ in the offset models converged to the fiducial value within 25\% of the statistical precision.

\subsection{Checks for instrumental and astrophysical systematic effects}\label{ss:consistency}

Further to the main contributors to the systematic error budget assessed above (\Sref{ss:syserr}), we considered several other possible effects below. We do not find evidence that these contribute to the systematic error budget, so we list them only as upper limits in \Tref{t:results}. Nevertheless, they are important for demonstrating the robustness of the \daa\ measurement from this single absorber.

\subsubsection{LFC versus ThFP wavelength calibration}\label{sss:LFCvsThFP}

In principle, the laser frequency comb (LFC) should provide a much more reliable wavelength calibration for ESPRESSO than the combined ThAr$+$Fabry--P\'erot (ThFP) approach. Each LFC mode's wavelength is known, in absolute terms, with $\ll$1\,\cms\ accuracy, while ThAr laboratory wavelengths typically have $\approx$2--20\,\ms\ uncertainties \citep{Redman_2014ApJS..211....4R} and, potentially, systematic errors that vary with wavelength. The number density of LFC modes also far exceeds that of ThAr lines, and adjacent modes typically have similar intensity, unlike the ThAr lines. However, instrumental effects that differ between the quasar and LFC exposure can still lead to systematic errors in the quasar wavelength scale. In particular, the quasar and calibration light will illuminate the optical fibre differently and, despite significant effort to ensure high optical scrambling efficiency (e.g.\ octagonal fibres and a double scrambler unit), their point-spread functions (PSFs) on the detector may differ. More importantly, any PSF difference may be a function of position on the detector, leading to intra-order and, possibly, longer-range distortions of the quasar relative wavelength scale and, therefore, systematic effects in \daa.

While we have no direct means to gauge this effect, some insight is gained from \citet{Schmidt_2021A&A...646A.144S}'s direct comparison of ESPRESSO's LFC and ThFP calibrations. Their figure 15 demonstrates that the two wavelength solutions differ by 0--20\,\ms, with $\sim$5\,\ms modulation within the red arm's echelle orders (up to $\sim$15\,\ms\ in the blue) and larger variations over longer wavelength scales. While the latter could plausibly be due to systematic errors in the ThAr laboratory wavelengths, the former must be due to instrumental effects that differ for the LFC and ThAr light, such as the differential PSF variations described above. It is therefore possible that similar systematics affect the quasar wavelength scale when calibrated with the LFC, as in our quasar exposures.

As a guide to the likely size of this effect on \daa, we re-measured it using ThFP-calibrated quasar exposures. Instead of the LFC calibration, we derived the wavelength solution for each quasar exposure from the standard ThFP calibration taken on the same day. The extracted quasar exposures were combined in exactly the same way as before, and the fiducial profile model for the right region was re-optimised on the new spectrum. The ThFP-calibrated \daa\ differed from the LFC-calibrated value by only 0.02\,ppm. This clearly demonstrates that our \daa\ value is robust to systematic effects in the wavelength calibration of ESPRESSO. We emphasise that this result may not generalise to other ESPRESSO spectra or absorption systems (i.e.\ comparisons of different wavelength regions); future analyses of other absorbers should explicitly test ESPRESSO's wavelength scale for systematic errors.

\subsubsection{Combined LFC$+$ThFP wavelength calibration}\label{sss:combLFCThFP}

As described in \Sref{ss:wavecal}, the LFC wavelength calibration only applies to the $\approx$4850--7000\,\AA\ range, while outside this range the wavelength scale is derived from the combined ThAr $+$ Fabry--P\'erot (ThFP) daytime calibration. The \tran{Mg}{i}{2026}, \ion{Zn}{ii} and \ion{Cr}{ii} transitions all fall bluewards of the LFC-calibrated range (4357--4445\,\AA). If the ThFP calibration differs substantially from the LFC one, it is possible that these transitions will be spuriously shifted relative to the other, redder transitions fitted in the right region. To test for this effect, we introduced a further free parameter -- a velocity shift -- for each fitting region associated with these six bluest transitions. These velocity shifts will be almost entirely degenerate with the constraints on \daa\ from these transitions, while still allowing them to help constrain the velocity structure. We find a negligible change in \daa\ when introducing these additional free parameters: \daa\ increased by just 0.01\,ppm relative to the fiducial value in the right region. This is expected because (i) these six weak transitions do not strongly constrain \daa\ in this absorber; and (ii) the LFC--ThFP calibration difference is likely very small, $\sim$10\,\ms\ \citep{Schmidt_2021A&A...646A.144S}.

\subsubsection{Shifts between ESPRESSO arms}\label{sss:arms}

The blue and red arms of ESPRESSO are physically separated, with different collimators, cross-dispersers, cameras and detectors. A dichroic produces a small overlap in wavelength coverage between the two arms, 5191--5272\,\AA. In the $\zab=1.1508$ absorber, this means most transitions fall in the red arm, with \tran{Fe}{ii}{2382} and all bluer transitions in the blue arm. To test for any difference in calibration between the two arms, we again introduce a velocity shift parameter for each fitting region of the bluer transitions, as above for the LFC-versus-ThFP test. This means that only the red-arm transitions constrain \daa, while all transitions still constrain the velocity structure like in the fiducial fit. We find only a very small, 0.14\,ppm change in \daa\ when introducing these changes to the fit of the right region, with a slightly larger statistical uncertainty of $\sigma_{\rm stat}=1.6$\,ppm. That is, the change in \daa\ is $\sim$0.1$\sigma_{\rm stat}$, providing no indication for a systematic difference in wavelength calibration between the two ESPRESSO arms. The right region will be most sensitive to any such effect because it has the best statistical precision and includes more transitions in the blue arm, so we do not extend this test to the other regions.

\subsubsection{Shifts between traces}\label{sss:traces}

The two traces of light from each fibre on the CCD are produced by an anamorphic pupil slicer, which is the first element encountered by light emerging from the optical fibres within the spectrograph's vacuum enclosure \citep{Pepe_2021A&A...645A..96P}. In comparing the LFC and ThFP calibration solutions (see \Sref{sss:LFCvsThFP} above), \citet{Schmidt_2021A&A...646A.144S} found than one trace had considerably larger deviations between the solutions than the other. The effect was much larger in the blue arm, 0--20\,\ms, particularly towards the red side of echelle orders, while it remained below $\sim$3\,\ms\ throughout the red arm. Thus, there appears to be illumination-dependent systematic errors that may have had an impact on our \daa\ measurements. We tested for this effect by producing a `sub-spectrum' of each trace with {\sc uves\_popler}: after both traces were combined to form the main spectrum, we recombined spectra from each trace without modifying any clipped data regions, atmospheric line masking, or continuum fits. See \citet{Murphy_2019MNRAS.482.3458M} for details of this process. The fiducial model of the right region was fitted to the two trace sub-spectra simultaneously, but with free velocity shift parameters allocated to each fitting region of one trace; \daa\ was therefore determined entirely by the other trace. Two separate \daa\ values were measured in this way, one for each trace, but they differed by only 0.31\,ppm with 1.59\,ppm statistical uncertainties each (i.e.\ only a 0.14$\sigma$ difference). The systematic error in \daa\ for the right region is therefore $\la$0.16\,ppm due to shifts between traces.

\subsubsection{Velocity shifts between transitions}\label{sss:shifts}

One of the main assumptions in the MM method for measuring \daa\ in quasar absorbers is that the different ionic species share the same velocity structure, i.e.\ that there are no intrinsic velocity shifts between transitions of different species. While this assumption has been questioned \citep[e.g.][]{Levshakov_2005A&A...434..827L}, it appears unlikely that different ions would really be physically separated in individual clouds and, therefore, have velocities differing by $\sim$20\,\ms, corresponding to $\sim$1\,ppm precision in \daa\ -- see discussion in \citet[][their section 4.3.2]{Kotus_2017MNRAS.464.3679K}. The difficulty in modelling the velocity structure of the $\zab=1.1508$ absorber is more likely to cause spurious shifts between transitions. For example, if too few velocity components have been fitted to a feature, this `underfitting' will likely introduce shifts, especially between different ionic species but also between transitions of the same species. Small, intrinsic inconsistencies between the absorption profiles of different transitions may also arise if some of the absorbing gas only partially covers the background quasar, or if HE 0515$-$4414 is gravitationally lensed, with multiple, as-yet-unresolved sight-lines combining to produce the observed spectrum. Previous attempts to assess these effects in slit-based spectrographs \citep[e.g.][]{Kotus_2017MNRAS.464.3679K} have been hampered by both the long-range and intra-order distortions of the wavelength scale (see \Sref{s:intro}). Clearly, these will both shift different transitions relative to each other depending on where they fall on the CCD mosaic. However, ESPRESSO's high-accuracy wavelength calibration avoids this. Therefore, ESPRESSO offers an important, alternative check on profile modelling errors to those derived above (\Sref{sss:profile}).

\begin{figure}
\begin{center}
\includegraphics[width=0.98\columnwidth]{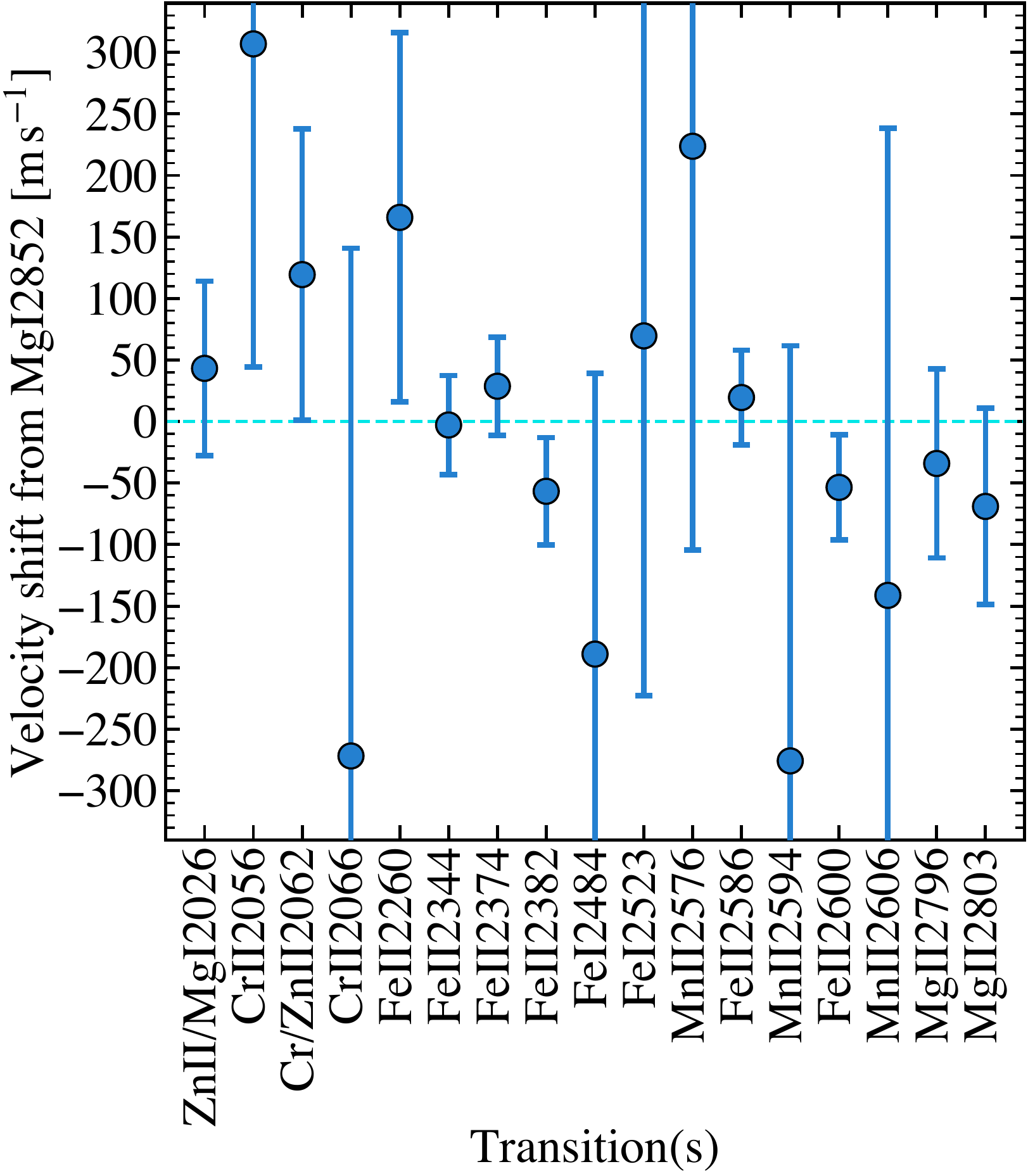} 
\vspace{-0.5em}
\caption{Velocity shifts between transitions in the right region, measured relative to \tran{Mg}{i}{2852}. \daa\ was fixed to zero while a separate velocity shift parameter was introduced for each fitting region in the fiducial fit (Figs.\ \ref{f:fit_r_strong} and \ref{f:fit_r_weak}). The best-fit values are shown together with their 1$\sigma$ statistical uncertainties. As illustrated in \Fref{f:fit_r_weak}, the \ion{Mg}{i} and \tran{Zn}{ii}{2026} transitions share the same fitting region, as do \ion{Cr}{ii} and \tran{Zn}{ii}{2062}, so they also share a common velocity shift measurement.}
\label{f:relshifts}
\end{center}
\end{figure}

\Fref{f:relshifts} shows the best-fitted velocity shifts between transitions for the right region. Here, \daa\ was fixed to zero in the fiducial model and a velocity shift was introduced as a free parameter for all fitting regions except that for \tran{Mg}{i}{2852}, i.e.\ all velocity shifts are measured with respect to this transition. \Fref{f:relshifts} clearly demonstrates that there are no systematic velocity shifts between transitions: all transitions are consistent with \tran{Mg}{i}{2852} and with each other. The stronger, unsaturated transitions in the right region also provide very precise velocity shift measurements: e.g.\ the five strong \ion{Fe}{ii} lines only have $\approx30$\,\ms\ uncertainties. Collectively, there is also no evidence for a shift between species: the weighed mean shift for the \ion{Fe}{ii} lines is $-6 \pm 16$\,\ms\ from the \tran{Mg}{i}{2852} line. Taken together, the results in \Fref{f:relshifts} support the assumptions of the MM method and indicate that our profile fitting approach is robust.

\subsubsection{Isotopic abundance variations}\label{sss:isotopes}

Most of the transitions fitted in the $\zab=1.1508$ absorber have isotopic structures whose components span velocity ranges of $\approx$0.4\,\kms\ for \ion{Fe}{ii} to $\approx$0.5--1\,\kms\ for the \ion{Mg}{i}/{\sc ii} transitions. As explained in \Sref{ss:fitting}, we assume the terrestrial isotopic abundances for each element in our analysis. However, if the abundances are different in the $\zab=1.1508$ absorber, we will spuriously infer shifts between different transitions and, therefore, in \daa. This is a well-recognised problem \citep{Murphy_2001MNRAS.327.1208M,Ashenfelter_2004PhRvL..92d1102A,Ashenfelter_2004ApJ...615...82A,Levshakov_2005A&A...434..827L} but, unfortunately, there are currently no strong and reliable constraints on the isotopic abundances in quasar absorbers \citep[e.g.][]{Agafonova_2011A&A...529A..28A,Levshakov_2009A&A...507..209L,Webb_2014MmSAI..85...57W,Vangioni_2019MNRAS.484.3561V,Noterdaeme_2018A&A...612A..58N}. For the $\zab=1.1508$ absorber, the $\approx$0.5\,\kms-wide isotopic structure of \tran{Mg}{i}{2852} is particularly important. If, for example, the dominant $^{24}$Mg isotope (79\% terrestrial abundance) was not present in the absorber and the two heavier isotopes ($^{25,26}$Mg) were equally abundant, this would cause a $\approx$6\,ppm spurious shift in \daa\ \citep[e.g.][]{Fenner_2005MNRAS.358..468F}. Possible isotopic abundance variations are therefore a potentially important systematic effect for our \daa\ measurements.

The velocity shift results in \Fref{f:relshifts} demonstrate that large isotopic abundance variations in the $\zab=1.1508$ absorber are very unlikely. The \ion{Fe}{ii} lines are shifted by just $-6 \pm 16$\,\ms\ relative to the \tran{Mg}{i}{2852} line in the right region (see \Sref{sss:shifts} above). For comparison, the velocity between the $^{24}\tran{Mg}{i}{2852}$ laboratory wavelength and the composite value with terrestrial isotopic abundances is 65\,\ms. So, if the heavier isotopes ($^{25,26}$Mg) have zero abundance in the absorber, we should see a velocity shift of this order between the \ion{Fe}{ii} transitions and \tran{Mg}{i}{2852} (assuming \daa\ is zero). Similarly, if only the heavier isotopes were present, with equal abundances in the absorber, we would expect to measure a 241\,\ms\ shift between the \ion{Fe}{ii} and \tran{Mg}{i}{2852} lines. Clearly, these extreme isotopic abundance variations are not favoured by the results in \Fref{f:relshifts}. However, a robust estimate of the isotopic abundance of Mg in the $\zab=1.1508$ absorber is not possible with such a simplistic analysis, and we leave this for future work. Nevertheless, given the difference of $\approx$0.04 in $Q$ between the \ion{Fe}{ii} and \tran{Mg}{i}{2852} transitions (see \Fref{f:Qvswl}), the observed shift of $-6 \pm 16$\,\ms\ between them corresponds to a systematic error in \daa\ of $<$0.7\,ppm.

With their high \SN\ UVES spectrum, \citet{Kotus_2017MNRAS.464.3679K} were able to strongly constrain the effect of isotopic abundance variations on their measurement of \daa\ in the same $\zab=1.1508$ absorber. They removed the influence of \ion{Mg}{i} and {\sc ii} transitions on \daa\ by introducing velocity shift parameters (as we did for several tests in this Section), finding a negligible 0.24\,ppm change in \daa, with only a marginal increase in the uncertainty. The latter point is due to the availability of the \tran{Fe}{ii}{1608} transition in the UVES spectrum: its large \emph{negative} $Q$ coefficient offers a high contrast with the other \ion{Fe}{ii} transitions which all have large \emph{positive} $Q$ values. We cannot perform the same test with our ESPRESSO spectrum because it does not cover the \tran{Fe}{ii}{1608} transition (which falls at $\approx$3460\,\AA, cf.\ ESPRESSO's blue wavelength limit of $\approx$3800\,\AA). Nevertheless, the consistency between our \daa\ measurement, which depends on the Mg isotopic abundances, and \citet{Kotus_2017MNRAS.464.3679K}'s which does not (see \ref{ss:previous} below) indicates that the Mg isotopic abundance effect is likely smaller than $\approx$1\,ppm in our measurement, consistent with the more direct estimate above.

We include our systematic error upper limit of $<$0.7\,ppm in \Tref{t:results}. As in previous studies \citep[e.g.][]{Kotus_2017MNRAS.464.3679K,Milakovic_2021MNRAS.500....1M}, we do not include this astrophysical effect in the formal systematic error budget for \daa. However, it is possible that the ESPRESSO data presented here, and the existing UVES spectrum, could be analysed together to formally, and more robustly, constrain (or perhaps measure) the $^{25,26}$Mg/$^{24}$Mg abundance ratio directly. This would then provide a more stringent measurement or limit on the effect on \daa\ in this specific absorber. We defer this to future work.

\subsection{Combined result for the entire absorber}\label{ss:comb}

\Tref{t:results} summarises the fiducial fitting results with the 1$\sigma$ statistical and systematic uncertainties for each of the three regions. The \daa\ values and uncertainties for each region are independent, so straight-forward Gaussian error propagation can be used to compute the weighted mean result for the entire absorber:
\begin{equation}\label{e:result}
\daa = 1.3 \pm 1.3_{\rm stat} \pm 0.4_{\rm sys}\,{\rm ppm},
\end{equation}
with 1$\sigma$ statistical and systematic error components. The right region dominates this combined result. In particular, the three main systematic effects (\Sref{sss:redisp}--\ref{sss:convergence}) cause similar uncertainties in \daa\ for the right region, and so make similar contributions to the systematic error combined across the regions.

The result in \Eref{e:result} is consistent with no change in the fine-structure constant between the absorber at $\zab=1.1508$ and the current laboratory value. The systematic uncertainty is well below the statistical one, which is remarkable given that the latter is similar to the ensemble precision obtained from previous large samples of absorbers (see \Sref{ss:results} for discussion). This is because ESPRESSO's accurate LFC wavelength calibration removes the systematic effects that undermine confidence in those large-sample results. Even the more recent results which were corrected for these effects using asteroid and/or solar twin spectra had residual wavelength calibration uncertainties 0.5--3.5\,ppm, larger than our total systematic error budget \citep[e.g.][]{Evans_2014MNRAS.445..128E,Murphy_2017MNRAS.471.4930M}. As explained in \Sref{ss:syserr}, \citet{Kotus_2017MNRAS.464.3679K} corrected the wavelength scale of their UVES spectrum of HE\,0515$-$4414 by using the HARPS spectrum of the same quasar. However, this procedure still left a $\sim$0.6\,ppm wavelength calibration uncertainty which dominated their systematic error budget. The recent \daa\ measurement in the $\zab=1.1508$ absorber by \citet{Milakovic_2021MNRAS.500....1M} used an LFC-calibrated HARPS spectrum to measure \daa, thereby avoiding wavelength calibration uncertainties like our ESPRESSO spectrum. Of course, the lower \SN\ of the HARPS spectrum ($\approx$58 per \kms, cf.\ 170 for our ESPRESSO spectrum) limited the precision to 2.4\,ppm (see further discussion in \Sref{ss:previous} below).

\section{Discussion}

\subsection{No substantial wavelength calibration errors}\label{ss:wavecal_disc}

As noted in \Sref{s:intro}, several of ESPRESSO's design features should strongly suppress systematic errors in \daa\ measurements -- i.e.\ relative velocity shifts between transitions at different wavelengths. However, even though the wavelength scale in this study was calibrated with the laser frequency comb (LFC), instrumental effects -- particularly illumination-dependent PSF variations across the spectrum -- may still affect measurements of \daa\ in general \citep{Schmidt_2021A&A...646A.144S}. The four tests described in Sections \ref{sss:LFCvsThFP}--\ref{sss:traces} all probe these possible wavelength calibration errors. The results are summarised in the lower section of \Tref{t:results}. For the specific case of the $\zab=1.1508$ absorber studied here, the effect on any wavelength calibration errors on \daa\ is negligible. However, we again emphasise that this may not be true for ESPRESSO studies of other absorbers, particularly those where important transitions fall towards the red edges of echelle orders in the blue arm. We recommend that quasar observations with ESPRESSO be complemented with asteroid or solar twin exposures to 
check the wavelength scale, especially if the LFC is not operational \citep[e.g.][]{Molaro_2008A&A...481..559M,Rahmani_2013MNRAS.435..861R,Evans_2014MNRAS.445..128E,Whitmore_2015MNRAS.447..446W}.

\subsection{Comparison with previous measurements in the $\zab=1.1508$ absorber}\label{ss:previous}

\begin{figure}
\begin{center}
\includegraphics[width=0.98\columnwidth]{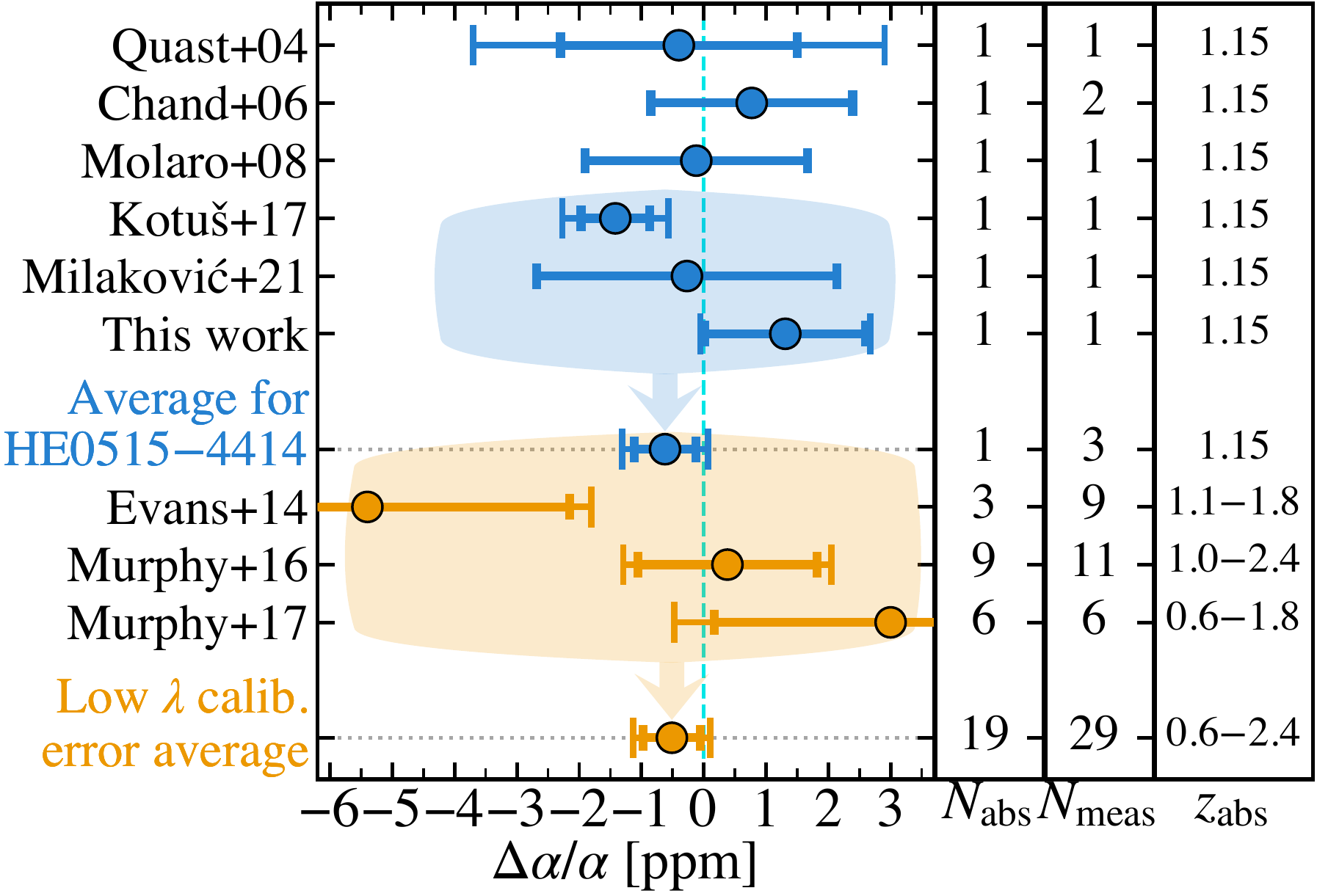} 
\vspace{-0.5em}
\caption{Measurements of \daa\ in the $\zab=1.1508$ absorber towards HE\,0515$-$4414 (blue points), and recent measurements which were corrected for, or insensitive to, wavelength calibration errors (orange points). The shaded patches show the values included in the weighted mean values indicated by arrows and dotted lines. For each study, the number of absorption systems, $N_{\rm abs}$, number of independent \daa\ measurements (for the entire absorber), $N_{\rm meas}$, and absorption redshift or range, $\zab$, are shown in the right-hand panels. Shorter, thicker error bars indicate 1$\sigma$ statistical uncertainties while the longer error bars with taller terminators show the quadrature sum of statistical and systematic uncertainties. The individual studies are discussed in Sections \ref{ss:previous} and \ref{ss:otherabs}.}
\label{f:all_da}
\end{center}
\end{figure}

The previous measurements of \daa\ at $\zab=1.1508$ towards HE\,0515$-$4414 are compared to our new value from ESPRESSO in \Fref{f:all_da} (blue points). All measurements are consistent with no variation in $\alpha$, and are also consistent with our result. The early UVES measurements by \citet{Quast_2004A&A...415L...7Q}, \citet{Chand_2006A&A...451...45C} and \citet{Molaro_2008EPJST.163..173M} \citep[who revised their earlier measurement in][]{Levshakov_2006A&A...449..879L} will all be subject to additional systematic errors, to differing degrees, from the long-range wavelength calibration distortions discussed in \Sref{s:intro} \citep[e.g.][]{Rahmani_2013MNRAS.435..861R,Whitmore_2015MNRAS.447..446W}. \citet{Kotus_2017MNRAS.464.3679K} demonstrated this explicitly, finding that these earlier results most likely require corrections by $\sim+$2\,ppm, similar to their statistical uncertainties. \citeauthor{Kotus_2017MNRAS.464.3679K} corrected for these effects in their \daa\ measurement by directly comparing their UVES spectra to a (ThAr-calibrated) HARPS spectrum of the same quasar. The HARPS ThAr calibration has no evident long-range distortions \citep{Whitmore_2015MNRAS.447..446W}. \citet{Milakovic_2021MNRAS.500....1M} performed the first LFC-calibrated measurement of \daa\ with new HARPS observations, with a separate study of wavelength calibration errors demonstrating that they are negligible for that work \citep{Milakovic_2020MNRAS.493.3997M}.

Focussing on these three recent, independent results with relatively low wavelength calibration errors, \Fref{f:all_da} indicates the formal weighted mean value for the $\zab=1.1508$ absorber towards HE\,0515$-$4414:
\begin{equation}\label{e:HE0515av}
\daa = -0.62 \pm 0.50_{\rm stat} \pm 0.48_{\rm sys}\,{\rm ppm}.
\end{equation}
Again, this is consistent with no variation in $\alpha$. It is also consistent, at 1.6$\sigma$, with the expected value from the $\alpha$ dipole of \citet{King_2012MNRAS.422.3370K}, i.e.\ $2.2\pm1.6$\,ppm. This value derives from the simplest dipole model where $\daa = 10.2_{-1.9}^{+2.2}\cos(\Theta)$\,ppm for $\Theta$ the angle between HE\,0515$-$4414 and the model's pole (${\rm RA}=17.4\pm0.9$\,h, ${\rm Dec.}=-58\pm9^\circ$). The line of sight towards HE\,0515$-$4414's is almost orthogonal to the dipole direction ($\Theta=78^\circ$), so the model expectation is near zero, i.e.\ HE\,0515$-$4414 does not provide a strong test of the dipole model.

We note that the weighted mean in \Eref{e:HE0515av} is dominated by the \citet{Kotus_2017MNRAS.464.3679K} measurement from wavelength-corrected UVES spectra. Despite their considerable efforts to estimate the error budget correctly, it remains plausible that their statistical and systematic uncertainties may be underestimated. For example, the model non-uniqueness problem discussed below (\Sref{ss:structure}) could affect both the statistical uncertainty and contribute a larger systematic error than previously estimated. These problems potentially extend to all measurements, but it is prudent to be cautious when combining results, especially when one dominates. If we exclude that measurement, the weighted mean of \citet{Milakovic_2021MNRAS.500....1M}'s HARPS result and our ESPRESSO measurement is $\daa = 0.93 \pm 1.14_{\rm stat} \pm 0.33_{\rm sys}\,{\rm ppm}$, where the statistical uncertainty is, of course, larger than in \Eref{e:HE0515av} but the systematic uncertainty is smaller, reflecting the suppression of systematic effects in the two LFC-calibrated measurements.

\subsection{Comparison with recent measurements in other absorbers}\label{ss:otherabs}

\Fref{f:all_da} also compares our result, and the average for the $\zab=1.1508$ absorber [\Eref{e:HE0515av}], to other recent measurements which had relatively small wavelength calibration uncertainties (orange points). \citet{Evans_2014MNRAS.445..128E} used asteroid and iodine-cell exposures to correct the wavelength scales of UVES, HIRES and Subaru/HDS spectra of the same quasar. Similarly, \citet{Murphy_2017MNRAS.471.4930M} corrected HDS spectra of two quasars using exposures of solar twins. Using a different approach, \citet{Murphy_2016MNRAS.461.2461M} measured \daa\ using only the (normally weak) Zn/\ion{Cr}{ii} transitions in `metal-strong' absorbers towards 9 quasars. These ions' transitions have large sensitivity coefficients ($Q$) with opposite sign, making them sensitive probes of $\alpha$ but, most importantly, the transitions have very similar wavelengths, rendering the \daa\ measurements insensitive to long-range calibration distortions. Combining the 28 results of these studies with the weighted mean for $\zab=1.1508$ towards HE\,0515$-$4414 [\Eref{e:HE0515av}], we obtain
\begin{equation}\label{e:all_da_av}
\daa = -0.51 \pm 0.48_{\rm stat} \pm 0.42_{\rm sys}\,{\rm ppm}.
\end{equation}
This weighted mean result represents the most reliable measurement using quasar absorption lines, and is relatively free of the long-range wavelength calibration errors that affected all previous spectra, including the large samples from HIRES \citep{Webb_2001PhRvL..87i1301W,Murphy_2003MNRAS.345..609M,Murphy_2004LNP...648..131M} and UVES \citep{Webb_2011PhRvL.107s1101W,King_2012MNRAS.422.3370K}. As with \Eref{e:HE0515av}, the \citet{Kotus_2017MNRAS.464.3679K} measurement in the $\zab=1.1508$ absorber also dominates \Eref{e:all_da_av}. Excluding that single result yields a weighted mean of $\daa = 0.51 \pm 0.82_{\rm stat} \pm 0.37_{\rm sys}\,{\rm ppm}$, which still represents ppm-level precision. Again, as in \Sref{ss:previous}, we urge some caution when combining many results together: if uncorrected systematic errors remain, or some measurements have underestimated total uncertainties, these will significantly affect the weighted mean result.

For 27 of the 29 measurements with low wavelength calibration errors in \Fref{f:all_da} -- those prior to 2021 -- \citet{Martins_2017PhRvD..95b3008M} and \citet{Murphy_2017MNRAS.471.4930M} have already noted that they do not support the $\alpha$ dipole model. Ignoring uncertainties in the model itself, those 27 results are inconsistent with it at the $\approx$4$\sigma$ level. However, if the dominant \citet{Kotus_2017MNRAS.464.3679K} measurement is removed, this falls to $\approx$2.5$\sigma$ \citep{Murphy_2017MNRAS.471.4930M}. Our new ESPRESSO measurement, and \citet{Milakovic_2021MNRAS.500....1M}'s HARPS measurement, do not appreciably alter this conclusion, given the near orthogonal directions to HE\,0515$-$4414 and the dipole.

\subsection{Velocity structure}\label{ss:structure}

\begin{figure}
\begin{center}
\includegraphics[width=0.98\columnwidth]{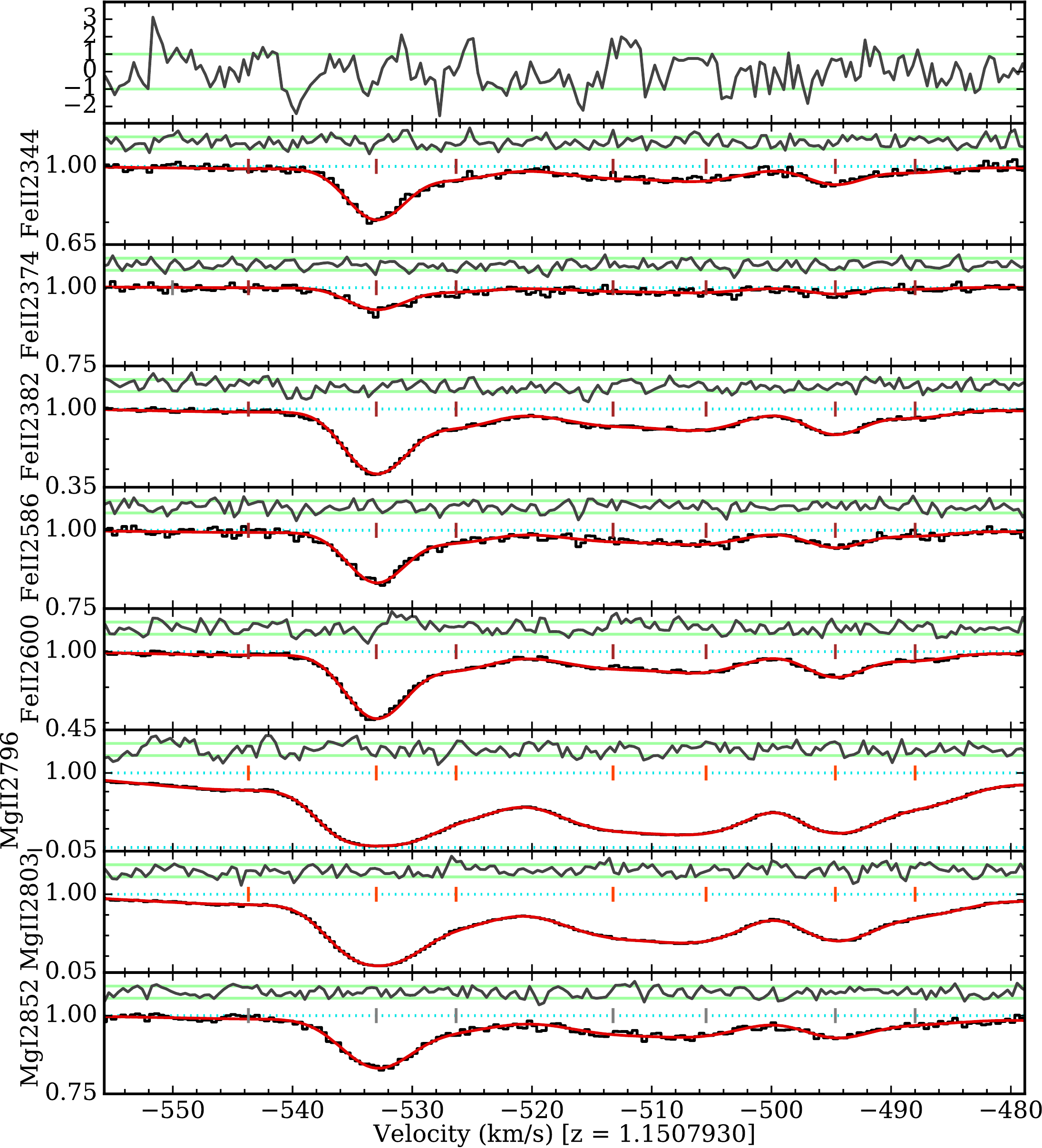} 
\vspace{-0.5em}
\caption{Same as \Fref{f:fit_l} but using the velocity structure of \citet{Milakovic_2021MNRAS.500....1M} for their region I (corresponding to part of our left region) which used a compound broadening model (i.e.\ turbulent and thermal broadening mechanisms). This model was used to fit our ESPRESSO spectrum with {\sc vpfit}. Note the significant, correlated structures in the composite residual spectrum (top panel) and in individual transitions, especially \tran{Mg}{ii}{2796} and \tran{Fe}{ii}{2600} near the strongest absorption feature.}
\label{f:fit_l_comp}
\end{center}
\end{figure}

As shown in \Tref{t:fit_comp}, our fitting approach requires a large number of velocity components to fit all the statistical structure of the ESPRESSO spectrum: 48 in the right region of the $\zab=1.1508$ absorber, compared to 49 in \citet{Kotus_2017MNRAS.464.3679K}'s fit to their UVES spectrum, and 26 for \citet{Milakovic_2021MNRAS.500....1M}'s HARPS spectrum. We mentioned in \Sref{ss:model_right} that the former two are likely similar because the ESPRESSO and UVES spectra have similar \SN\ per \kms\ and because the fitting approach was very similar. The HARPS spectrum's \SN\ per \kms\ is much lower than ESPRESSO's ($\approx$58 vs.\ $\approx$170), so we should expect fewer components to be fitted by \citet{Milakovic_2021MNRAS.500....1M}. However, they also used a very different fitting approach: an automated algorithm for determining the velocity structure with {\sc vpfit}, called {\sc ai-vpfit} \citep{Lee_2021MNRAS.504.1787L}, and both turbulent and thermal broadening for each component's $b$ parameter (`compound' broadening). As a simple illustration, \Fref{f:fit_l_comp} shows the compound model of \citet{Milakovic_2021MNRAS.500....1M} for part of their region I (corresponding to part of our left region), with $\chisq$ minimised using {\sc vpfit} on our ESPRESSO spectrum. It is immediately clear that this fit does not adequately account for the statistical structure in the data: e.g.\ the composite residual spectrum shows several systematic excursions, correlated over many pixels, particularly around the strongest absorption at $-540$ to $-525$\,\kms, but at other velocities as well. This clearly demonstrates that the larger \SN\ per \kms\ of the ESPRESSO spectrum drives the need for a more complex fit; differences in the fitting approach cannot fully explain this.

\citet{Lee_2021MNRAS.507...27L} explored how the {\sc ai-vpfit} approach produced different velocity structures from different starting points of the algorithm and different broadening model assumptions, and how \daa\ varied between these models. A similar `non-uniqueness' problem arises for the more interactive, human-guided approach we have employed: different people will choose to introduce new components at different velocities, with different initial parameters, while building up a complex fit, and so produce somewhat different velocity structures for the same dataset. We have attempted to ensure our measurement is robust to this effect by thoroughly exploring alternative velocity structures and incorporating their variance in \daa\ into our error budget -- see \Sref{sss:profile}. However, we limited that search to adding or subtracting one component from our fiducial model, as the sheer number of other possibilities is not possible to explore in this way. \citet{Lee_2021MNRAS.507...27L} find that turbulent-only models within {\sc ai-vpfit} produce a larger variety of fits, with a larger range of \daa\ values than the compound broadening approach. While this is certainly a concern, it is important to recognise that this is unlikely to apply simplistically outside {\sc ai-vpfit} itself. Our turbulent-only fitting approach maximises the number of components using $\chisq_\nu$ as the information criterion, while the addition of new components -- where in the profile they are added, and with what initial parameters -- is judged by the human user. Importantly, that judgement seeks to avoid leaving inconsistencies in the model between different ions and/or transitions. Currently, {\sc ai-vpfit} appears to solve such potential inconsistencies, at least sometimes, by adding blending components (i.e.\ from an unidentified species in an unidentified absorber) in some fitting regions; these random blends differ when the algorithm is started with different random seeds -- see figures A1--A3 of \citet{Lee_2021MNRAS.507...27L} which cover the same fitting range as \Fref{f:fit_l_comp}. A full {\sc ai-vpfit} analysis of our ESPRESSO spectrum would be very welcome: we make our ESPRESSO spectrum publicly available in \citet{Murphy_2021Spec+Fits_HE0515} online for others to analyse, or we will undertake this if {\sc ai-vpfit} becomes available.

\section{Conclusions}

In this work we have made the first precise measurement of \daa\ with the new ESPRESSO spectrograph on the VLT. HE\,0515$-$4414 was observed for 16.1\,h to achieve a very high \SN\ ($\approx$105 per 0.4\,\kms\ pixel at 6000\,\AA) at resolving power $R \approx 145000$, with the wavelength scale of all 17 exposures calibrated using the laser frequency comb (LFC). The entire analysis procedure was developed using a blinded version of the spectrum to avoid human biases; once the blinding was removed, the analysis was re-run without alteration or human interaction. The high $R$ and \SN\ revealed that the $\zab=1.1508$ absorption system was more complex than previous studies found: strong constraints on the relative optical depths of two different \ion{Mg}{i} lines confirmed the presence of very narrow velocity components ($b<0.5$\,\kms) in the strongest absorption feature  (\Sref{ss:model_right}). A total of 129 velocity components were required to fit the $\approx$720\,\kms-wide absorption profile, with the strongest \ion{Mg}{i} and \ion{Fe}{ii} transitions providing the main constraints on \daa.

Our new measurement at $\zab=1.1508$ from the ESPRESSO spectrum is
\begin{equation}\label{e:da_conclusion}
\daa = 1.3 \pm 1.3_{\rm stat} \pm 0.4_{\rm sys}\,{\rm ppm}\,,
\end{equation}
consistent with no cosmological variation in the fine-structure constant. The total error is similar to the ensemble precision of the previous, large samples of absorbers from Keck/HIRES and VLT/UVES that indicated variations at the $\approx$5\,ppm level \citep{Webb_2001PhRvL..87i1301W,Murphy_2003MNRAS.345..609M,Murphy_2004LNP...648..131M,Webb_2011PhRvL.107s1101W,King_2012MNRAS.422.3370K}, which likely arose from long-range distortions in the wavelength scale \citep[e.g.][]{Rahmani_2013MNRAS.435..861R,Whitmore_2015MNRAS.447..446W}. Our measurement is free from those systematic errors due to the specific design features of ESPRESSO to suppress them (e.g.\ octagonal fibre feed, stable vacuum environment, and LFC calibration). However, we emphasise that wavelength calibration errors are still possible with ESPRESSO, and future studies of \daa\ should specifically test for them, especially when important transitions fall in the blue arm (\Sref{ss:wavecal_disc}). The main systematic uncertainties in our measurement arise from ambiguities in fitting the absorption profile, effects from redispersion of the spectra, and convergence of the fitting procedure (\Sref{ss:syserr}).

HE\,0515$-$4414 is very bright (\textsl{Gaia} $G=14.9$\,mag), which has enabled many previous measurements of \daa\ in the $\zab=1.1508$ absorber. Our new \daa\ measurement in \Eref{e:da_conclusion} is consistent with the two most recent of those from two different spectrographs: the first LFC-calibrated measurement, from HARPS \citep[][$\daa=-0.3\pm2.4$\,ppm]{Milakovic_2021MNRAS.500....1M}, and the most precise measurement in a single absorber, from a very high \SN\ UVES spectrum \citep[][$\daa=-1.4\pm0.6_{\rm stat}\pm0.6_{\rm sys}$\,ppm]{Kotus_2017MNRAS.464.3679K}. Both are relatively free from wavelength calibration uncertainties (see \Sref{ss:previous}) and can be combined with our ESPRESSO measurement -- see[\Eref{e:HE0515av}]:
\begin{equation*}
\daa = -0.6 \pm 0.5_{\rm stat} \pm 0.5_{\rm sys}\,{\rm ppm}.
\end{equation*} 
A sample of 26 other measurements with HIRES, UVES and Subaru/HDS which have been corrected for, or are insensitive to, long-range wavelength distortions are also consistent with our new measurement and have a similar overall precision: $\daa = 0.4 \pm 1.3_{\rm stat} \pm 0.7_{\rm sys}$\,ppm \citep{Evans_2014MNRAS.445..128E,Murphy_2016MNRAS.461.2461M,Murphy_2017MNRAS.471.4930M}. Combining the 29 independent, low-wavelength calibration error measurements of \daa\ at $\zab=0.6$--2.4 above yields a weighted mean [\Eref{e:all_da_av}]:
\begin{equation*}
\daa = -0.5 \pm 0.5_{\rm stat} \pm 0.4_{\rm sys}\,{\rm ppm}\,.
\end{equation*}
\Fref{f:all_da} summarises these comparisons with previous measurements.

ESPRESSO promises to provide a larger sample of well-calibrated, high quality quasar absorption spectra for measuring \daa\ through the instrument consortium's Guaranteed Time Observations and open, competitive observing time. One important, outstanding problem in improving quasar absorption measurements of \daa\ is the lack of observational constraints on how the isotopic abundances in the absorbers differ from the terrestrial values. The overall velocity shifts between \tran{Mg}{i}{2852} and the strong \ion{Fe}{ii} lines in our ESPRESSO spectrum limit the effect on our measurement of \daa\ to $<$0.7\,ppm (\Sref{sss:isotopes}). It may be possible to directly measure the Mg isotopic abundance ratio in this absorber with our spectrum, given ESPRESSO's excellent wavelength calibration accuracy and, in principle at least, detailed map of the instrumental line-shape function provided by the LFC and Fabry--P\'erot calibration exposures. However, given its complex absorption profile, a robust isotopic ratio estimate in the $\zab=1.1508$ absorber will likely require an even more extensive exploration of alternative velocity structures than we have performed here.
%This would be a natural application of the {\sc ai-vpfit} approach of \citet{Lee_2021MNRAS.504.1787L}, as would a reassessment of the value and uncertainties in \daa\ from our ESPRESSO spectrum.

\begin{acknowledgements}

We dedicate this paper to the memory of Dieter Reimers, discoverer of quasar HE\,0515$-$4414, leader in European astronomy, colleague and friend to many authors of this paper, who passed away in June 2021.
The authors acknowledge the ESPRESSO project team for its effort and dedication in building the ESPRESSO instrument.
We thank the anonymous referee for their careful appraisal of the draft manuscript and comments that improved it.
MTM acknowledges the support of the Australian Research Council through Future Fellowship grant FT180100194 and the Institute for Fundamental Physics of the Universe for hosting a collaborative visit in June 2019.
The INAF authors acknowledge financial support of the Italian Ministry of Education, University, and Research with PRIN 201278X4FL and the ``Progetti Premiali'' funding scheme.
TMS also acknowledges the support from the Swiss National Science Foundation (SNSF) and the University of Geneva.
This work was supported by FCT -- Funda\c c\~ao para a Ci\^encia e a Tecnologia through national funds and by FEDER through COMPETE2020 -- Programa Operacional Competitividade e Internacionaliza\c c\~ao by these grants:
UID/FIS/04434/2019;
UIDB/04434/2020;
UIDP/04434/2020;
PTDC/FIS-AST/32113/2017 \& POCI-01-0145-FEDER-032113;
PTDC/FIS-AST/28953/2017 \& POCI-01-0145-FEDER-028953;
PTDC/FIS-AST/28987/2017 \& POCI-01-0145-FEDER-028987;
CERN/FIS-PAR/0037/2019 and PTDC/FIS-OUT/29048/2017.
This project has received funding from the European Research Council (ERC) under the European Union’s Horizon 2020 research and innovation programme (project {\sc Four Aces}; grant agreement No 724427). It has also been carried out in the frame of the National Centre for Competence in Research PlanetS supported by the Swiss National Science Foundation (SNSF).
M-RZO acknowledges funding under project PID2019-109522GB-C51 of the Spanish Ministerio de Ciencia e Investigaci\'on.
VA acknowledges support from FCT through Investigador FCT contract number IF/00650/2015/CP1273/CT0001.
RR, CAP, JIGH, and ASM acknowledge financial support from the Spanish Ministry of Science and Innovation (MICINN) project AYA2017-86389-P. JIGH also acknowledges financial support from the Spanish MICINN under 2013 Ram\'on y Cajal program RYC-2013-14875.
DE acknowledges financial support from the Swiss National Science Foundation for project 200021\_200726.
This research made use of {\sc astropy} \citep{Astropy_2013A&A...558A..33A,Astropy_2018AJ....156..123A}; and {\sc matplotlib} \citep{Hunter_2007CSE.....9...90H}.

\end{acknowledgements}

%\bibliographystyle{aa} 
%\bibliography{references}

\begin{appendix}

\section{Alternative profile models}\label{s:models}

Here we detail the alternative models used to derive an estimate of the systematic errors from profile model non-uniqueness in \Sref{sss:profile}. \Tref{t:models} identifies the velocity at which components were removed from, or added to the fiducial models for the right, central and left regions plotted in Figs.\ \ref{f:fit_r_strong}, \ref{f:fit_c} and \ref{f:fit_l} respectively. In general, we focused on components in the spectral features that most strongly influence \daa. We also investigated other features where two components were fitted close to each other -- we removed one of them to assess whether it was really statistically required. The stability, or otherwise, of each model is noted in the table. Note that \Tref{t:models} does not show all models attempted: many models were initially investigated which, upon preliminary analysis, were either highly unstable or produced substantially larger \chisq\ per degree of freedom ($\chisq_\nu$) than the fiducial model and/or large deviations in the composite residual spectra; these were discarded without further consideration.

\begin{table}
\caption{Alternative profile models attempted in each region. The velocity at which a component was added ($+$) or removed ($-$) is measured relative to the absorption redshift, $\zab=1.150793$. A model is deemed unstable if {\sc vpfit} rejected important components in one species that were relatively strong in other species. For stable fits, the final two columns provide the change in \daa\ and the Akaike information criterion, corrected for finite sample sizes, AICC, relative to the fiducial model in Figs.\ \ref{f:fit_r_strong}, \ref{f:fit_c} and \ref{f:fit_l}.}
\label{t:models}
\vspace{-1.5em}
\begin{center}
\begin{tabular}{lccccc}
\hline\hline
Velocity & Added or    & Stable? & $\Delta(\daa)$ & $\Delta\chi^2$ & $\Delta(\textrm{AICC})$ \\
($\kms$) & removed?    &         & (ppm)          &                &                         \\
\hline
\multicolumn{5}{l}{\textbf{Right region}} \\
$ -38$ & $+        $ & Yes & $-0.24$ & $ -8.8 $ & $ -0.1 $ \\
$ -24$ & $+^{\rm a}$ & Yes & $ 0.02$ & $  4.3 $ & $  6.5 $ \\
$  -3$ & $-        $ & Yes & $-2.28$ & $ 53.4 $ & $ 31.5 $ \\
$  6 $ & $-        $ & No  &         &          &          \\
$  14$ & $+        $ & Yes & $ 0.26$ & $ -0.7 $ & $ 12.4 $ \\
$  16$ & $-        $ & No  &         &          &          \\
$  18$ & $+        $ & Yes & $ 0.31$ & $ -0.8 $ & $  7.9 $ \\
$  29$ & $+        $ & Yes & $-0.15$ & $  0.0 $ & $ 11.0 $ \\
$  52$ & $-        $ & No  &         &          &          \\
$  55$ & $-        $ & No  &         &          &          \\
$  60$ & $-        $ & Yes & $ 0.04$ & $ 15.7 $ & $  4.7 $ \\
$  76$ & $-        $ & Yes & $ 0.42$ & $ 15.7 $ & $  0.4 $ \\
\multicolumn{5}{l}{\textbf{Central region}} \\
$-321$ & $-$ & Yes & $ 0.70$ & $11.9$ & $ 5.5$ \\
$-244$ & $-$ & Yes & $ 0.07$ & $ 2.6$ & $ 2.6$ \\
$-238$ & $-$ & No  &         &        &        \\
$-232$ & $-$ & Yes & $ 5.06$ & $15.9$ & $ 5.1$ \\
$-219$ & $+$ & Yes & $-1.26$ & $-9.3$ & $ 5.8$ \\
$-218$ & $-$ & Yes & $-0.11$ & $10.7$ & $ 8.7$ \\
$-179$ & $-$ & No  &         &        &        \\
$-127$ & $-$ & Yes & $-2.00$ & $ 2.1$ & $-6.6$ \\
$-123$ & $-$ & No  &         &        &        \\
\multicolumn{5}{l}{\textbf{Left region}} \\
$-543$ & $-$ & Yes & $ 0.02$ & $ 10.4$ & $  3.8$ \\
$-536$ & $-$ & Yes & $-0.37$ & $ 23.6$ & $ 12.5$ \\
$-523$ & $-$ & Yes & $-0.20$ & $  9.6$ & $ -1.4$ \\
$-513$ & $-$ & Yes & $ 0.10$ & $ 20.0$ & $  6.8$ \\
$-507$ & $-$ & Yes & $-0.07$ & $ 12.8$ & $  1.8$ \\
$-492$ & $-$ & Yes & $ 1.80$ & $  5.2$ & $-10.2$ \\
$-467$ & $-$ & No  &         &         &         \\
$-416$ & $-$ & Yes & $-2.99$ & $ 87.1$ & $ 76.1$ \\
$-391$ & $-$ & Yes & $ 1.58$ & $ 10.2$ & $  3.6$ \\
\hline
\end{tabular}
\end{center}
$^\textrm{a}$The component at $-24$\,\kms\ was only added to \ion{Fe}{ii} as it already exists in \ion{Mg}{i}/{\sc ii} in the fiducial model (see \Fref{f:fit_r_strong}). While this model is listed as stable here, this additional \ion{Fe}{ii} component's column density slowly reduces over repeated runs of {\sc vpfit}, with $\chisq$ barely changing, and it is rejected eventually.
\end{table}

For the remaining models, \daa\ was measured in the same way as for the fiducial model (\Sref{ss:fitting}), first using the blinded spectrum and again after unblinding (see \Sref{ss:blinding}). Initially, \daa\ was fixed at the fiducial value while the alternative model was constructed and refined by minimising \chisq\ in {\sc vpfit}. Then {\sc vpfit} was re-run with \daa\ as a free parameter. \Tref{t:models} shows how \daa\ deviates from the fiducial value for the region, and the increase in \chisq\ and the Akaike information criterion, corrected for finite sample sizes, AICC, as a result. Note that some stable alternative models have a lower AICC than the fiducial model. At first, it may seem that one of these models was, therefore, more appropriate to adopt as fiducial. However, this is not correct for two reasons: (i) the fiducial model was selected while the analysis was blinded, so the $\chisq_\nu$ and AICC values were based on the blinded spectrum, while those in \Tref{t:models} are from the final, unblinded analysis; and (ii) we used $\chisq_\nu$ as a simpler information criterion for deciding which model should be the fiducial one, while AICC is perhaps more justified for comparing very similar models as we do in this systematic error analysis.

The most important aspects of the alternative models are discussed below for each region.

\subsection{Right region}

The spectral features that most strongly influence \daa\ are those at 0 and 9\,\kms, and many components were already required for our fiducial model in these features, particularly the former (5 components). We already discussed in \Sref{ss:model_right} that no reasonable fit could be constructed without the very narrow `cold' components in the 0\,\kms\ feature. However, removing a single broader component in the 0\,\kms\ feature, like that at $-3$\,\kms, results in a much higher AICC. The only velocity at which a new component could be added to these features was at 14\,\kms\ which produced a somewhat higher AICC (12.4) but very little change in \daa\ (0.3\,ppm). In the rest of the profile, a 0.4\,ppm change in \daa\ resulted from removing one of the two components in the sharp right-hand edge of the 75\,\kms\ feature. This feature is saturated in \ion{Mg}{ii} and the two components in the fiducial fit were found to provide a marginally smaller AICC (by 0.4).

\subsection{Central region}

The structure of the most important spectral features for measuring \daa, at $-237$ and $-140$ to $-120$\,\kms, were thoroughly explored in this region. In the former, removing one of the reddest 3 components leaves a stable, statistically acceptable fit with an AICC `only' 5 more than the fiducial fit. This fit leads to quite a large change in \daa, 5\,ppm, approximately the same as the statistical uncertainty in this region. While the AICC change of 5 is within our accepted AICC change threshold of 10, it is still regarded as `strong' evidence against this model \citep{Jeffreys_1961}. Nevertheless, we include it in our estimate for profile modelling systematic errors in this region. For the double feature between $-140$ and $-120$\,\kms, removing the component at $-127$\,\kms\ in \ion{Mg}{i}/{\sc ii} results in a lower AICC, but \daa\ only changes by 2\,ppm. Although the feature at $-179$\,\kms\ is relatively broad and does not constrain \daa\ strongly, it is interesting that removing one of the two components at its centre resulted in an unstable fit, causing other surrounding components to be rejected by {\sc vpfit} in some species.

\subsection{Left region}

\daa\ is mainly constrained by the features at $-533$, $-495$ and $-470$\,\kms, and the three between $-420$ and $-380$\,\kms. \Tref{t:models} shows that the velocity structure fitted to the strongest of these, at $-533$\,\kms, seems to have little effect on \daa, while that fitted to the $-495$ and $-390$\,\kms\ features is less well determined -- shifts in \daa\ of $\approx$1.5--2\,ppm are found for alternative models which have similar AICC values as the fiducial model. We found no alternative stable fits with more components than the fiducial model.

\end{appendix}

\end{document}